\documentclass{article}
\usepackage{multicol}
\usepackage{float} 
\usepackage{PRIMEarxiv}
\usepackage{dblfloatfix}
\usepackage{duckuments}
\usepackage[utf8]{inputenc} 
\usepackage[T1]{fontenc}    
\usepackage[hidelinks]{hyperref} 
\usepackage{url}            
\usepackage{booktabs}       
\usepackage{amssymb} 
\usepackage{amsfonts}       
\usepackage{amsmath}
\newcommand{\R}{\ensuremath{\mathbb{R}}}
\usepackage{nicefrac}       
\usepackage{physics}
\usepackage{microtype}      
\usepackage{lipsum}
\usepackage{fancyhdr}       
\usepackage{graphicx,caption}   
\usepackage{svg}
\usepackage{adjustbox}
\graphicspath{{media/}}     
\usepackage[backend=biber,style=apa,natbib=true]{biblatex}
\usepackage{authblk}

\usepackage{hyperref}

\hypersetup{
    colorlinks=true,
    linkcolor=blue,
    filecolor=magenta,      
    urlcolor=cyan, 
    citecolor=cyan,
    }

\title{Spiking neurons as predictive controllers of linear systems}

\author[1]{Paolo Agliati}
\author[2]{André Urbano}
\author[2]{Pablo Lanillos}
\author[1]{Nasir Ahmad}
\author[1]{Marcel van Gerven}
\author[1]{Sander Keemink}

\affil[1]{Department of Machine Learning and Neural Computing, Donders Institute for Brain, Cognition and Behaviour, Radboud University, Nijmegen, the Netherlands}
\affil[2]{Cajal International Neuroscience Center, Spanish National Research Council, Madrid, Spain}

\begin{document}
\maketitle

\begin{abstract}
Neurons communicate with downstream systems via sparse and incredibly brief electrical pulses, or spikes. Using these events, they control various targets such as neuromuscular units, neurosecretory systems, and other neurons in connected circuits. This gave rise to the idea of spiking neurons as controllers, in which spikes are the control signal. Using instantaneous events directly as the control inputs, also called `impulse control', is challenging as it does not scale well to larger networks and has low analytical tractability. Therefore, current spiking control usually relies on filtering the spike signal to approximate analog control. This ultimately means spiking neural networks (SNNs) have to output a continuous control signal, necessitating continuous energy input into downstream systems. Here, we circumvent the need for rate-based representations, providing a scalable method for task-specific spiking control with sparse neural activity. In doing so, we take inspiration from both control theory and neuroscience, and define a spiking rule where spikes are only emitted if they bring a dynamical system closer to a target. From this principle, we derive the required connectivity for an SNN, and show that it can successfully control linear systems. We show that for physically constrained systems, predictive control is required, and the control signal ends up exploiting the passive dynamics of the downstream system to reach a target. Finally, we show that the paradigm scales to both high-dimensional systems and bio-inspired motor control tasks. Importantly, in all cases, we maintain a closed-form mathematical derivation of the network connectivity, the network dynamics and the control objective. This work advances the understanding of SNNs as biologically-inspired controllers, providing insight into how real neurons could exert control, and enabling applications in neuromorphic hardware design.
\end{abstract}
\vspace{1cm}

\begin{multicols}{2}

\section*{Introduction}

\begin{figure*} [!b]
\centering
\adjustbox{trim=0.1cm 0.05cm 0.05cm 0.05cm, clip}{
\includegraphics[width=\textwidth]{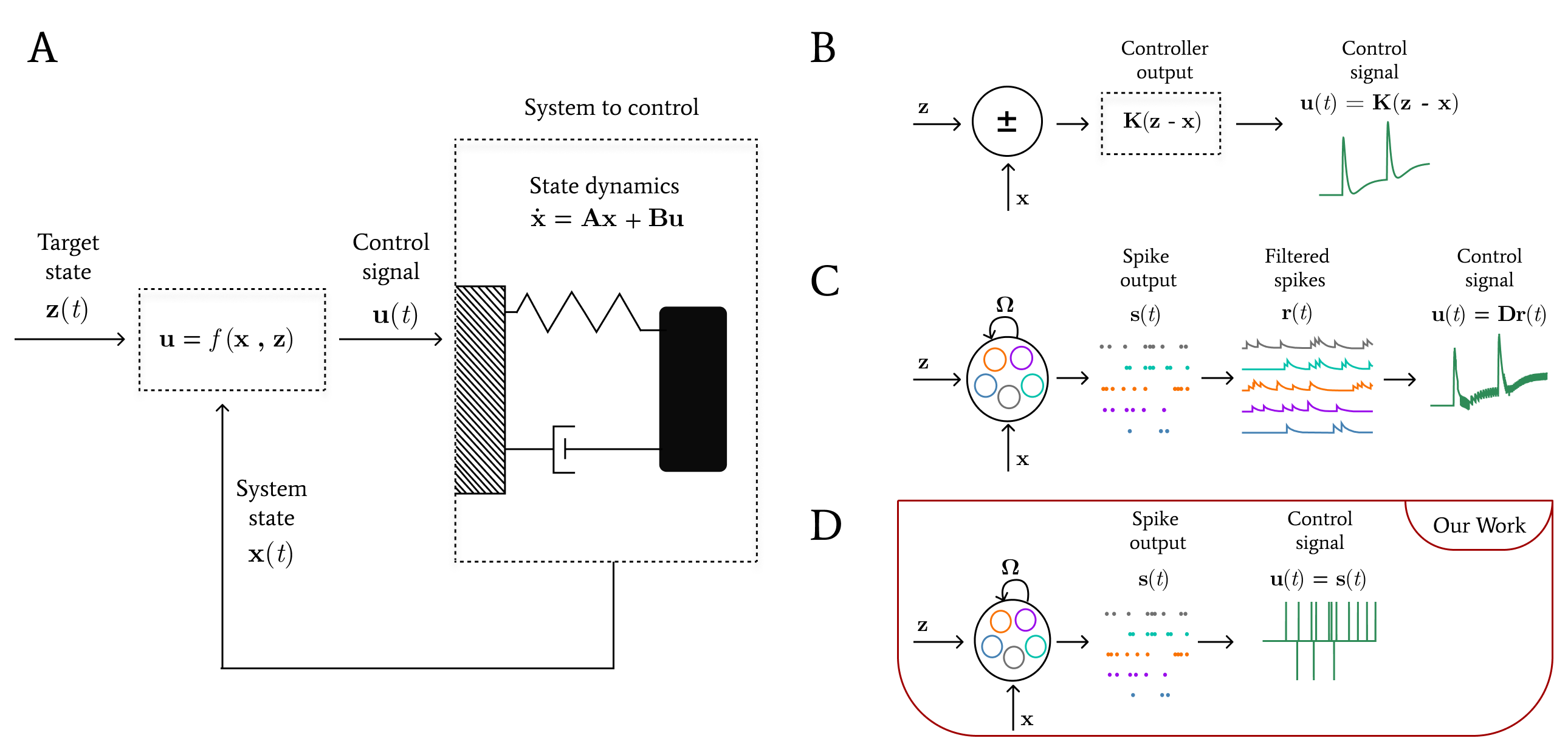}}
\caption{\textbf{Overview of feedback control approaches.}
A: Scheme of a feedback controller and a downstream system. The current state of the system $\mathbf{x}\mathit{(t)}$ is given as input to the controller, along with a target value $\mathbf{z}\mathit{(t)}$. The controller outputs a control signal $\mathbf{u}\mathit{(t)}$, as a function of both inputs. $\mathbf{u}\mathit{(t)}$ is then mapped onto the dynamics of the downstream system via the $\mathbf{B}$ matrix. The system's dynamics are governed by the $\mathbf{A}$ matrix. B-D: Examples of three feedback control paradigms. B: The continuous LQR uses a gain matrix $\mathbf{K}$ to directly compute the control signal for a given state and target pair. C: In the filtered spiking case, the output spikes of the network are filtered to obtain $\mathbf{r}\mathit{(t)}$, which is decoded by the matrix $\mathbf{D}$ to form the control signal $\mathbf{u}\mathit{(t)}$ (adapted from~\cite{slijkhuisClosedform2023}). D: In our spiking control paradigm, the output spikes $\mathbf{s}\mathit{(t)}$ represent the control input itself, and are therefore directly mapped onto the system dynamics through the $\mathbf{B}$ matrix.}
\label{fig1}
\end{figure*}

In the vertebrate nervous system, action potentials are one of the universal principles of communication between neurons~\cite{kochBiophysics2004, gollischRapid2008, wolfeSparse2010, boerlinSpikebased2011, gutigspike2014, srivastavaMotor2017}.
Spiking behavior of cortical neurons grants efficient and robust computational properties,  providing a basis for the realization of the cortex's complex adaptive functions~\cite{mainenReliability1995, kochBiophysics2004, bazhenovRole2005, gollischRapid2008, wolfeSparse2010, boerlinSpikebased2011}. At the level of single neurons, spike events are incredibly brief and sparse, and through synaptic communication can affect the dynamics of neurons in downstream areas. This property, where neurons are not only communicating information but modifying behavior, settled the basis for viewing neurons as controllers~\cite{cisekcomputer1999, ahissarPerception2016} and spikes as control signals. 
We can find clear examples of spike-driven control in movement control studies, where the spikes' influence allows animals to orient their head in space~\cite{zhangRepresentation1996, eliasmithunified2005}, as well as control the movement of their eyes~\cite{seungHow1996, eliasmithunified2005} or limbs~\cite{eliasmithunified2005, bernikernormative2019}.
This control perspective can offer important insights into neural computation, but is also increasingly relevant for neuromorphic applications where spiking networks are used to control downstream plants~\cite{stagstedneuromorphic2020, polykretisspiking2022, liuSpiking2023}.

Although the theory and applications of control with spiking networks has recently seen some advances, it remains unclear how to generate each spike to successfully control a plant's dynamics. Traditionally, control theory relies on optimal control methods, such as the linear quadratic regulator (LQR) ~\cite{andersonOptimal2007}, which continuously adjusts control inputs to drive a system toward a target state while minimizing a cost function (Fig.~\ref{fig1}B). Although effective, these continuous control paradigms are difficult to reconcile with the discrete nature of SNNs. Attempts at bridging this gap include the use of a filtered spikes approach. These studies have applied classic optimal control principles to SNNs by approximating LQR signals through rate codes or filtered spike trains~\cite{boerlinPredictive2013, deneveEfficient2016, huangDynamical2018, guoNeural2021, slijkhuisClosedform2023}, as shown in Fig.~\ref{fig1}C. Importantly, in these filtered spikes approaches, spikes serve as the network's representation of the system state but are not actively computing control signals, which are still continuous in nature. This makes it difficult to trace the effect each single spike has on the trajectory of the controlled system.
Alternative approaches for spike-based control in the literature focus on individual neuron-based controllers~\cite{mooreneuron2024}, or implement forms of instantaneous control~\cite{yangImpulsive1999}. Such impulsive control methods have been widely studied, but they mostly focus on the control theory perspective, sometimes even bypassing the need of a network~\cite{yangStability2007, raffObservers2007, wangImpulsive2014, ouyangImpulsive2020}. Only recently, some works started integrating this framework with theoretical neuroscience concepts, to explicitly interpret the control impulses as the neurons' action potentials~\cite{petriAnalysis2024}. Still, network architectures are imposed rather than derived from control objectives. For this reason, the integration of bio-informed SNNs with (feedback) control remains an under-explored area. A new paradigm is needed to account for direct spiking control and its seamless integration with computational neuroscience theories. 

Our study investigates how spike events can influence the dynamics of linear systems, using spikes directly as control signals. We consider this problem from the perspective of control theory, by formally deriving how neurons should spike to control a linear dynamical system (Fig.~\ref{fig1}).
This equates to generating control inputs that can successfully drive the dynamics of a downstream plant to minimize a certain loss associated with the state of the system.
When deriving a solution for this problem (see Materials and Methods), we obtain, without needing to introduce further assumptions, a feedback control scenario like the one depicted in Fig.~\ref{fig1}A. 
Our work embeds this instantaneous spiking control within theoretical neuroscience constraints.

Inspired by the neuroscience theory of spike coding networks (SCNs)~\cite{boerlinPredictive2013, gutigspike2014, deneveEfficient2016, guoNeural2021}, we derive, from control theory principles, a network of recurrent integrate-and-fire (IF) neurons with sparse activity and low-rank connectivity. By directly incorporating spikes as control signals (Fig.~\ref{fig1}D), our model ensures that each neuron's contribution to the system’s trajectory is explicit. Between each spike event $s(t)$, the system evolves following its dynamics in an open loop, and when a control signal is produced, the system switches to a closed-loop control (with $s(t)$ being the control input). Our solution naturally exploits the intrinsic dynamics of the plant, instead of continuously inputting a signal over time. This approach could help understanding how neural circuits can execute control tasks efficiently while preserving key features of biological networks. In the following sections, we present our framework, analyze the model's performance on a linear dynamical system, and discuss its implications for both neuroscience and control theory.

\section*{Results}

\subsection*{Formulating the control problem}
To investigate the role of spiking activity in controlling linear dynamical systems, we start by revisiting the fundamental components of a control task. We then introduce our approach to modeling spike-based control, and finally explore the consequences of applying this method to physically constrained systems. 

\subsubsection*{General control setup}
The control of a generic $K$-dimensional linear dynamical system through an $N$-dimensional control signal, as illustrated in Fig.~\ref{fig1}A,  can be expressed as:
\begin{equation}
\mathbf{\dot{x}} = \mathbf{A x} + \mathbf{B} \mathbf{u} \,,
\end{equation}
where $\mathbf{x} \in \mathbb{R}^K $ is the system state,
$\mathbf{A} \in \mathbb{R}^{K \times K} $ is the system matrix, describing the system's uncontrolled dynamics, $\mathbf{B} \in \mathbb{R}^{K \times N} $ is the input matrix which maps the control input onto the system, and $\mathbf{u} \in \mathbb{R}^N $ is the control input. 

In classical control, $\mathbf{u}$ is typically a continuous signal, as in LQR control, where the optimal control input is computed to minimize a quadratic cost function (Fig.~\ref{fig1}B). For control systems that make use of neural networks, the control signal can instead be derived from neural activity.
In spiking networks, a filtered-spike signal can be employed, where the firing rates of neurons determine the value of the control input for each timestep (Fig.~\ref{fig1}C). Here, the spikes are aiding the network's internal representation of the systems, but the control input still approximates a continuous control signal~\cite{boerlinPredictive2013, deneveEfficient2016, huangOptimizing2017, huangDynamical2018, guoNeural2021, slijkhuisClosedform2023, mooreneuron2024}.
In our work, we instead consider a sparse and discrete control signal, where the input is made of brief pulses or `kicks' (Fig.~\ref{fig1}D)~\cite{yangImpulsive1999, yangStability2007, raffObservers2007, wangImpulsive2014, ouyangImpulsive2020, petriAnalysis2024}. The system then becomes:
\begin{equation} \label{eq:dyn}
\mathbf{\dot{x}} = \mathbf{A x} + \mathbf{B} \mathbf{s} \,,
\end{equation}
where $\mathbf{s} \in \mathbb{R}^N$ describes the on/off state of each of $N$ neurons in the network. A spike signal is emitted every time the voltage $V_i$ of neuron $i$ reaches its threshold $T_i$, resulting in the spike train $s_i(t) = \sum_j \delta(t - t_j)$ with $t_j$ indexing the spike times.
Importantly, here $\delta$ is the Dirac delta function, which describes an instantaneous pulse (whose area integrates to one) at spike times $t_j$ and is zero otherwise. This is different from other forms of impulsive control like bang-bang control or chattering control~\cite{felgenhauerstability2003}, where the instantaneous input is maintained for a certain period of time. 

\subsubsection*{Spike-based control} 
To formalize our control approach, we define a loss function that expresses the discrepancy between the current state of the linear dynamical system and a desired target state, by taking the squared $L^2$ norm of their difference. Specifically, we consider the loss:

\begin{equation} \label{eq:control_loss}
L = \| \mathbf{x}(t+f) - \mathbf{z}(t+f) \|^2+ \mu \Gamma(\mathbf s) + \alpha ||\mathbf r ||^2 
\end{equation}
where $\mathbf{x}(t+f)$ is the system state some time $f$ in the future, $\mathbf{z}(t+f) \in \mathbb{R}^K$ the target state, $\mu$ is a regularization parameter that determines the spiking cost, $\Gamma (\mathbf s)= \int_{t}^{t+ \delta t} \mathbf s(t) \dd{t}$ counts the spikes within a short time window (in practice within a simulation timestep), $\mathbf s$ are the spike trains, $\alpha$ is a parameter that scales the cost for the neuron's activity, and  $\mathbf r$ are the neurons' filtered spike trains. We consider both reactive control ($f=0$) and predictive control ($f>0$).

Classically, such a loss might be minimized by taking the gradient with respect to certain system parameters, or by finding the optimal control input that minimizes this loss for some future time window, as in standard LQR control. However, the use of a spiking control signal makes both of these approaches challenging to solve mathematically. Here, we instead take inspiration from the neuroscientific theory of spike coding networks. SCN theory assumes spikes should only happen when they improve on an underlying coding loss and derives the required spiking dynamics. From this principle, spiking neural networks that represent their inputs~\cite{boerlinPredictive2013, deneveEfficient2016} and perform a variety of computations~\cite{brendelLearning2020, guoNeural2021, slijkhuisClosedform2023} can be defined.

\begin{figure} [H]
\adjustbox{trim=0.1cm 0.05cm 0.05cm 0.05cm, clip}{
\includegraphics[width=\columnwidth]{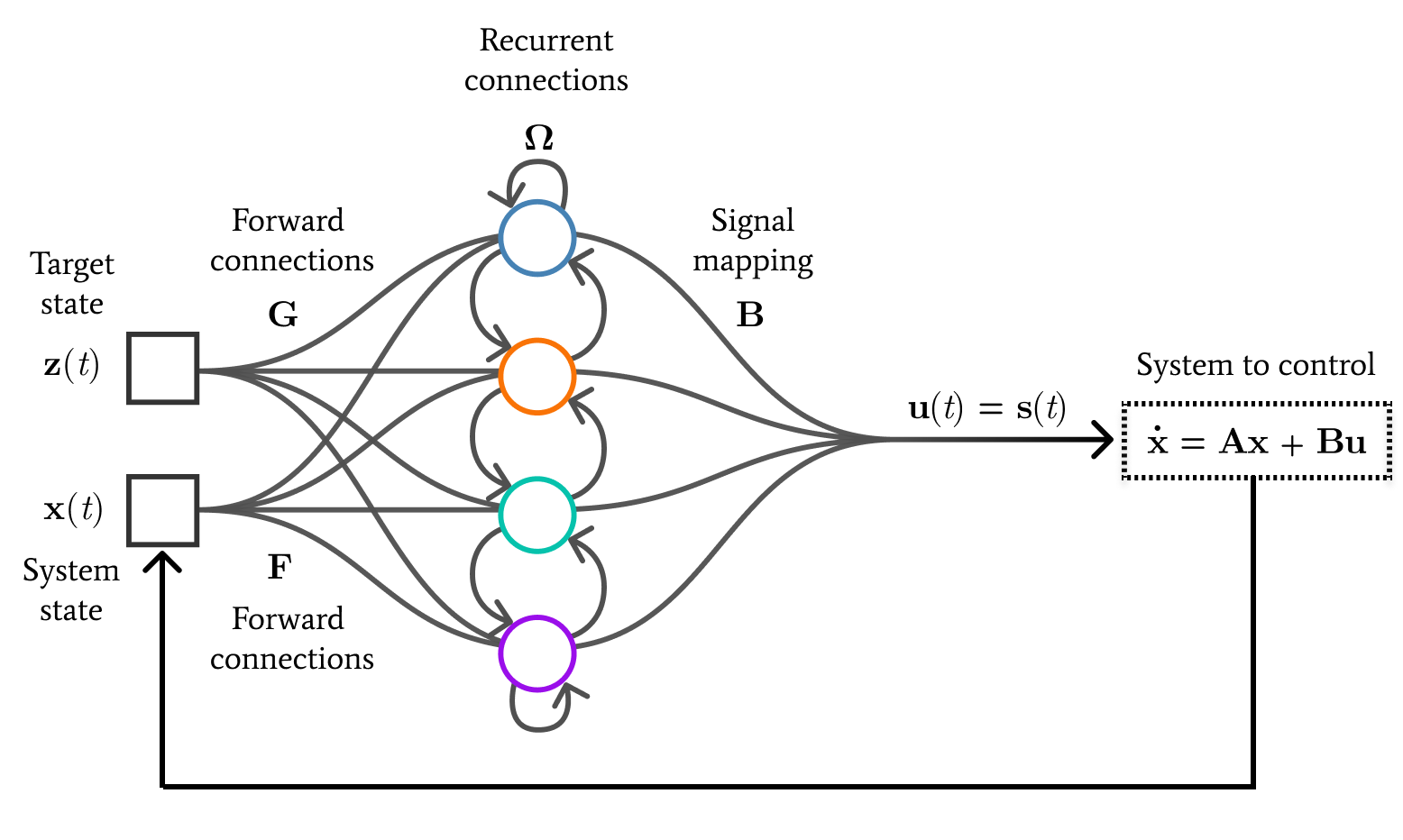}}
\caption{ \textbf{Illustration of the network-system loop.} 
Scheme of a recurrent spiking network composed of four neurons controlling a downstream system. Inputs and target are mapped onto the network via forward connections. $\mathbf{G}$ maps the target $\mathbf{z}$, while $\mathbf{F}$ maps the system's current state, $\mathbf{x}$. The matrix $\mathbf{\Omega}$ represents the recurrent weights. The control inputs are the spikes $\mathbf{s}$, which are mapped onto the system via the matrix $\mathbf{B}$.}
\label{fig2}
\end{figure}

Following this same principle, but applied to the problem of controlling dynamical systems, we require that a spike should only happen if it improves the control loss (Eq.~\eqref{eq:control_loss}). 
According to these SCNs-inspired approaches, whenever a neuron in the network spikes, it (directly or indirectly) influences the state of the plant, resulting in a different loss value. Therefore, we can compute two different losses $L^\text{s}_i$ and $L^\text{ns}$, representing the distance to the target in the presence or absence, respectively, of a spike from neuron $i$.
In order to control the plant's dynamics, the neurons in the network compare the two losses to determine whether inputting a spike is advantageous to reach the control target. 
The inequality $L^\text{s}_i < L^\text{ns}$ is describing how our loss can be improved by the control input, decreasing the distance towards the target and helping our network carry out the control task. Importantly, as described in the Discussion, the network will not compute nor approximate the optimal control solution for the problem. Differently from other methods (like the filtered spike controller illustrated in Materials and Methods), our spiking control algorithm does not track the LQR solution, but instead approaches the target by locally minimizing losses. 
These losses are describing a spiking condition, reminiscent of a voltage-threshold inequality $\mathbf{V} > \mathbf{T}$ (where $\mathbf{V}$ is the voltage of the neurons, and $\mathbf{T}$ is their firing threshold) that allows biological neural networks to produce action potentials. 
In Materials and Methods, we show that one can indeed derive voltage and thresholds values from our loss comparison. As a consequence, in our framework, the values of both voltage and threshold are composed of quantities that describe the (current and future) states of the plant. 
In particular, the voltage of neuron $i$ is expressed by:
\begin{equation} \label{eq:volnoc}
V_i = \mathbf{b}_i^T \mathbf{A}_f{^T} (\mathbf{\mathbf{z}} - \mathbf{A}_f \mathbf{x}_0) \, ,
\end{equation}
where $\mathbf{b}_i$ is the $i$th column of the matrix $\mathbf{B}$, $\mathbf{A}_f \equiv e^{\mathbf{A} f} \in \mathbb{R}^{K \times K}$ is the matrix exponential of the state matrix $\mathbf{A}$ times the future time window $f$, and $\mathbf{x}_0 = \mathbf{x}(t)$ is the current state of the system. The threshold of neuron $i$ is described as:

\begin{equation} \label{eq:thnoc}
T_i = \frac{\mathbf{b}_i^T \mathbf{A}_f{^T} (\mathbf{A}_f \mathbf{b}_i)  + \mu + \alpha (2\mathbf{r}_i + 1)} {2},
\end{equation} 
where $ \mathbf{r}_i$ indicates a vector containing only zeros except for the $i$th position, which holds the filtered spike traces of neuron $i$.
A detailed derivation of the values of $\mathbf{V}$ and $\mathbf{T}$ can be found in Materials and Methods. Although a cost $\mathbf{C}$ on each dimensions of the system can be introduced to further modulate spiking activity, the values of $V_i$ and $T_i$ in Eq.~\eqref{eq:volnoc} and Eq.~\eqref{eq:thnoc} do not consider this weighted case (assuming $\mathbf{C}$ is an identity matrix), for simplicity. Further developing this simple spiking rule leads to a recurrent network of integrate-and-fire neurons, illustrated in Fig.~\ref{fig2}, and described by:
\begin{equation}
\mathbf{\dot{V}} = -\mathbf{V} + \mathbf{G} (\dot{\mathbf{z}} + \mathbf{z}) - \mathbf{F} \mathbf{x}_0  -  \mathbf{\Omega s}\,,\end{equation}
where $\mathbf{V} \in \mathbb{R}^N$ represents the membrane potentials, $\mathbf{G} \in \R ^{N \times K}$ are the forward connections that encode the target $\mathbf{z}$, $\mathbf{F} \in \R ^{N \times K}$ contains the forward connections that encode the current system's state $\mathbf{x_0}$, $\mathbf{\Omega} \in \R ^{N \times N}$ represents recurrent connections, and $\mathbf{s}$ are the spikes. Interestingly, the obtained recurrent network has low-rank connectivity with respect to the dimensions of the downstream system (see Materials and Methods). This network structure is not assumed, but rather derived from each neuron as a greedy spiking condition given our loss comparison (see Discussion). Consequently, although the network obtained is recurrent, each neuron computes the losses without access to network-level information. For a full derivation of the network, see Materials and Methods. 

For this proof-of-concept, we enforce an asynchronous firing scheme: at each timestep, all neurons independently compute their local loss comparisons, but only one is allowed to spike, even if several exceed threshold. The spiking neuron can be chosen randomly from the suprathreshold set, or, as we do here, selected as the one with the highest voltage above threshold. While for clarity we keep this firing policy in all the shown examples, it is not strictly required for achieving control. This is because networks that operate on local spiking rules can distribute their activity very effectively among neurons~\cite{calaimgeometry2022}, which means no specific neuron is more relevant for the computation of the network. Such a simple asynchronous policy is only feasible in zero-delay systems, where the new state of the plant is inputted in the network without any time delay. In a system with delays, we could not impose this asynchronous firing. However, we can employ different methods that allow multiple neurons to spike while still being compatible with our local spiking rule, for instance, by scaling each neuron's threshold according to its past firing activity (see Materials and Methods). Examples of (failing and successful) spiking control of linear systems without the asynchronous firing constraint are shown in Supplementary Fig.~\ref{figA} in the Appendix.

\subsection*{Reactive vs. predictive spiking control} 

Our control problem is centered around the comparison of losses (distances of the state to a target) under the conditions in which a spike is generated or not (see Results). By comparing the loss functions in the spiking and non-spiking cases, we derived a firing condition for each neuron. Namely, the neuron's voltage exceeds its threshold and generates a spike only when the effect of this spike onto the system is to minimize the loss (reducing the distance between the system’s state and the target). Importantly, in our control paradigm, the spiking rule is fully local: each neuron computes its losses, with no access to shared network-level information, and fires only when the inequality $L^\text{s}_i < L^\text{ns}$ is satisfied, allowing the dynamics of the plant to unfold independently. 

In Results, we showed how starting with a spiking condition based on a loss comparison (spike if $L^\text{s}_i < L^\text{ns}$), one can derive the required network structure and dynamics to control a downstream system. Such a spiking rule can be defined relative to the immediate result of a spike, or the predicted effect that such spike will have on the system. In the first case, spikes are generated based on minimizing the loss function at the current time step. In the latter, spikes are generated based on a forward prediction of the system state, incorporating future expected loss minimization over a short horizon of $f$ seconds. The idea of predictive spiking is also present in SCN-related works, usually considering the spike's influence on a future state ~\cite{schwemmerConstructing2015}, or on the difference between past and current state ~\cite{zeldenrustEfficient2021}. Adapting this predictive spiking idea in our work, means that the system would evolve in open loop after the spike, only following its intrinsic dynamics, until another spike is able to further improve the predicted loss.

\begin{figure} [H]
\includegraphics[width=\columnwidth]{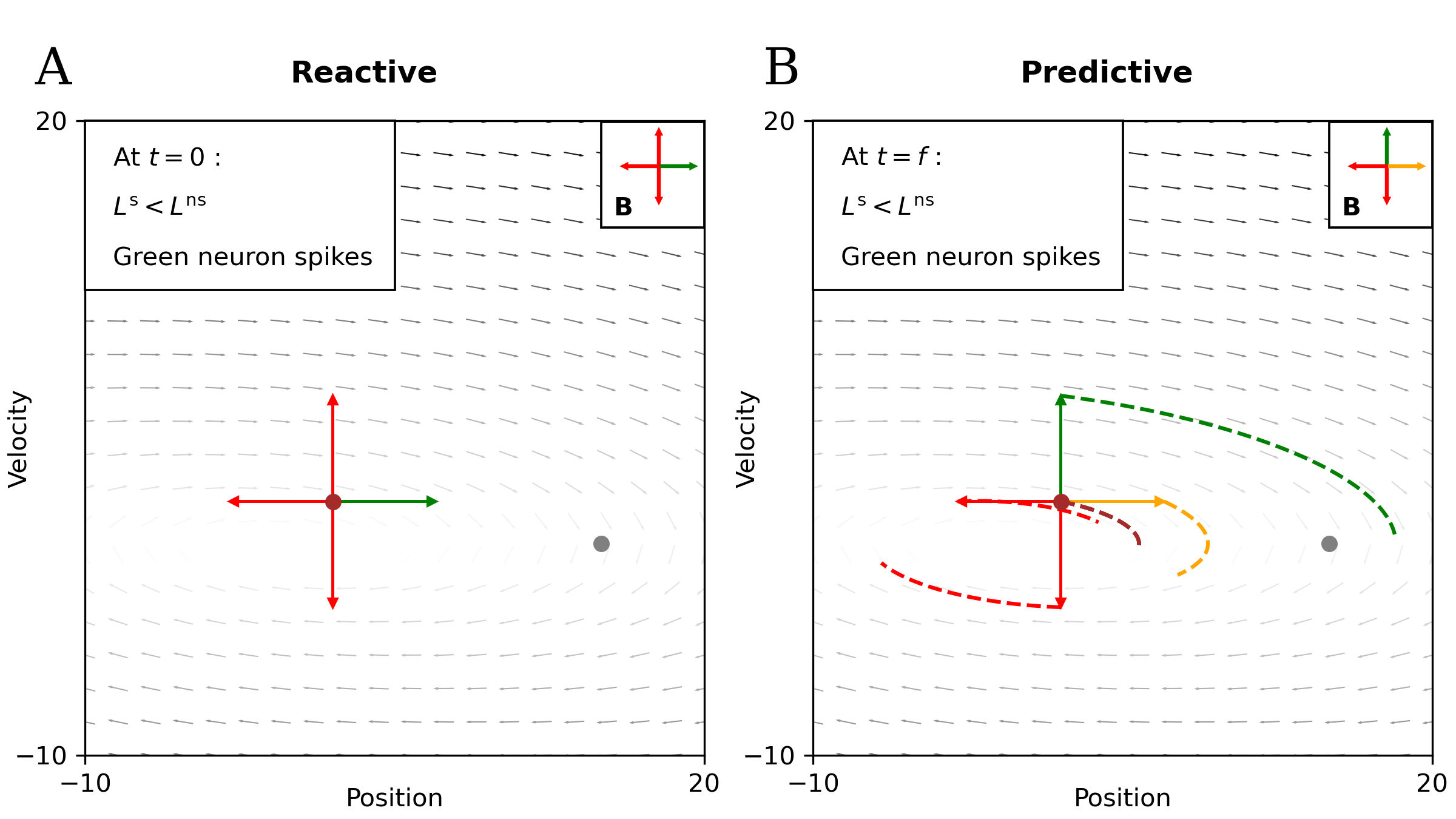}
\caption{ \textbf{Examples of reactive and predictive spiking.}
A: Example of a linear dynamical system with its current state (brown dot), and four example states that can be reached with a spike. The gray dot represents a target state. We employ a reactive spiking rule. The neuron for which $L^\text{s}_i < L^\text{ns}$ spikes (green arrow), since that immediately shortens the distance between the system's state and the target. The other ones do not (red arrows). 
B: Example of a linear dynamical system, where the spike decision is taken based on the state of the system $f$ seconds after the spike. In this predictive case, $L^\text{s}_i < L^\text{ns}$ is still the spiking condition, but the preferred spiking direction can differ. The condition $L^\text{s}_i<L^\text{ns}$ holds for both the upward (green) and rightward (orange) spikes, since the no-spike trajectory (brown) lies further from the target  (gray dot) than either of the spiking ones. Here, an asynchronous firing rule (see Results) would select only one of these neurons to fire; either at random, or by considering the neuron with the highest voltage (green).}
\label{fig3}
\end{figure}

Reactive and predictive spiking control could lead to significantly different spiking behavior. To illustrate this, we consider both cases for an example linear system, with its dynamics defined by velocity and position (Fig.~\ref{fig3}). Each spike produced by a neurons in the controlling network is multiplied by its corresponding column in the matrix $\mathbf B$, which maps it onto the dynamics of the system. In Fig.~\ref{fig3} we choose a $\mathbf B$ matrix that maps the spikes from four different neurons as vectors in the four different cardinal directions. When the system receives a spike control signal in the current timestep, the system state is instantaneously kicked in the corresponding direction. In the reactive case, only the immediate effect of the spiking is considered. This means the loss comparison that produces a spike control signal does not consider the future state of the plant, but rather aims to improve the loss right after the spike is emitted. As shown by Fig.~\ref{fig3}A, in this arbitrary linear system, adopting a reactive spiking rule results in a spike that kicks the system directly towards the target (green arrow), without accounting for the future dynamics of the plant (which could later move the trajectory further away from it). All other spiking directions (red arrows) do not result in an improved loss right after the spike, and so the neurons whose spikes are mapped in those directions do not fire.
On the contrary, predictive spiking control favors the spike that minimizes the loss $f$ seconds into the future, considering how the system would evolve during this time window. In our predictive example (Fig.~\ref{fig3}B), either spiking rightward (orange trajectory) or upward (green trajectory) would give a lower loss in $f$ seconds when compared to the non-spiking case (brown trajectory).
If we employ an asynchronous firing policy, only one of these two neurons is selected for firing. For these examples, we employ asynchronous firing and we select the neuron with the highest voltage past its threshold (the one whose spike is mapped on the green arrow). Note that our paradigm can work outside of this simple asynchronous firing case (Supplementary Fig.~\ref{figA} in the Appendix), by setting a spike-threshold adaptation mechanism (see Materials and Methods). 

\begin{figure*} [!t]
\includegraphics[width=\textwidth]{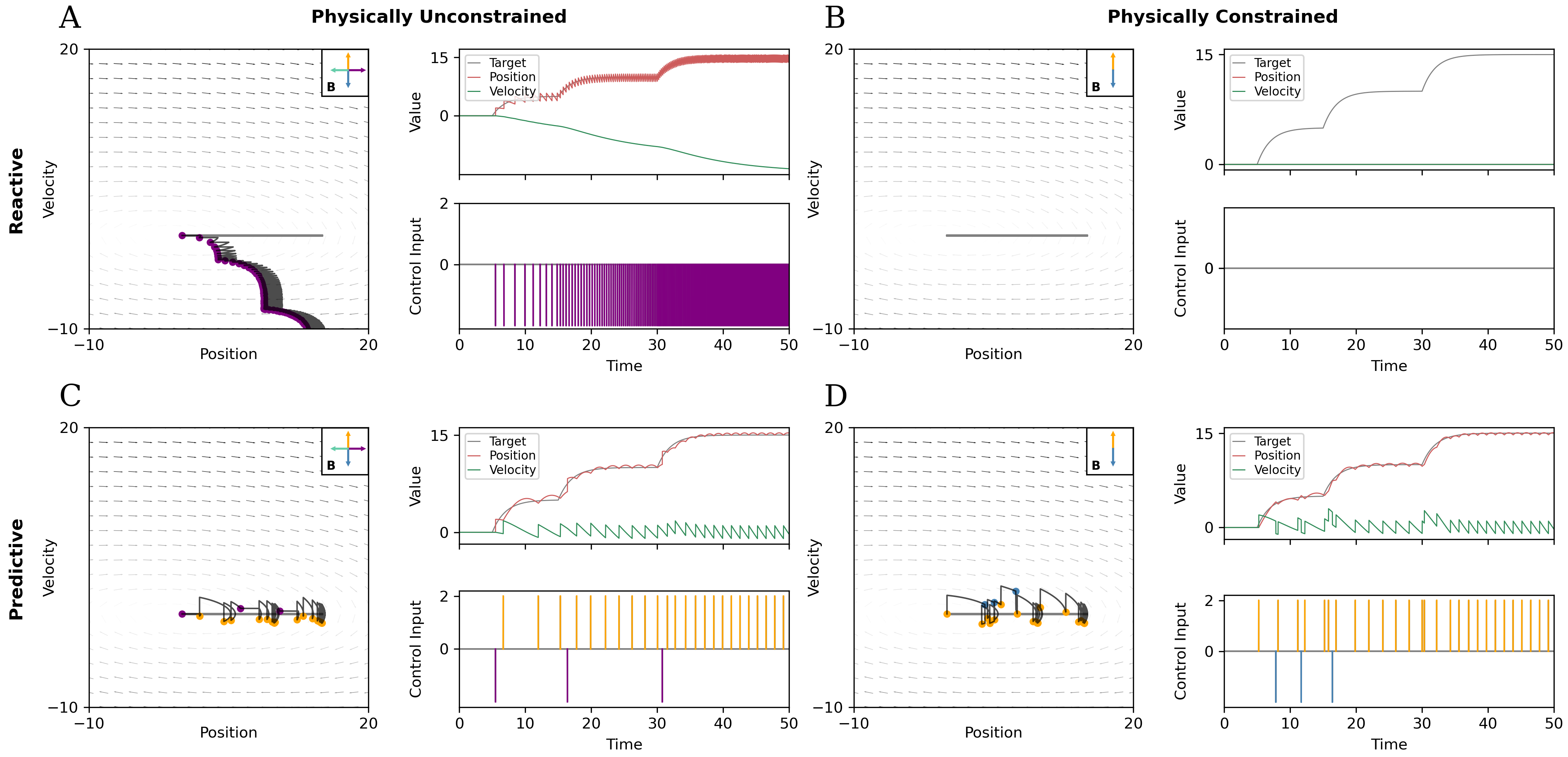}
\caption{\textbf{Spiking control simulations.}
The two rows denote reactive and predictive systems. The two columns represent a generic linear system, and the physical SMD. All the examples simulation are run with the simplest network possible of two neurons.
A: On the left side, state space representation of a reactive spiking control approach for a generic unconstrained system. The gray line corresponds to the changing target. Although the matrix $\mathbf{B}$ allows for spikes in all four directions, the network continuously spikes rightwards, ignoring the intrinsic dynamics of the plant. On the right side, the top figure shows the changes of velocity, position and target over time, while the bottom figure shows the spike control input (each spike scaled by corresponding value in the matrix $\mathbf{B}$).
B: On the left side, the state space figure shows that without predictiveness, a physically constrained system cannot be controlled using our greedy spiking rule. In this example, $\mathbf{B}$ is constrained to only allow spikes on the velocity component, which do not cause any immediate effect on the position, preventing the loss to be decreased by the spike events. When controlling a physical system with a reactive control approach, the loss in the non-spiking case is always smaller than the loss in the spiking case, and the network never spikes. On the right side, no activity is present. 
C: In the predictive, unconstrained example, the system can spike in all four directions but considers its future state. In this case, the network either spikes directly towards the target or spikes upwards, exploiting the systems dynamics.
D: Use of our predictive SNN approach for the control of an SMD. As shown in the state space representation, as the target position changes, the network spikes up or down, adjusting the velocity of the plant based on its predicted state in $f$ seconds. On the right side, we highlight how the network's activity adjusts to the target. It employs downwards spikes to adjust for an eventual overshooting of the position relative to the target, and produces consecutive upwards spikes when the target is changing more rapidly.}
\label{fig4}
\end{figure*}

\subsection*{Application to physical systems: spring-mass-damper}  
To illustrate an application of our framework, we consider, as an example, the control of a spring-mass-damper system (SMD)~\cite{guRobust2005}, for schematics see Fig.~\ref{fig1}. This is a linear dynamical system composed of a mass attached to a spring (which causes oscillations whenever the mass is moved) and a dampener (which dampens the oscillations, giving the system a fixed stable state). The system can be entirely described within its two variables (the position and the velocity of the mass) which are both included in our system state, $\mathbf{x}$.
This makes the SMD a very simple system that grants observable dynamics and clear analytical solutions, while also modeling a physical system with realistic state variables. As such, it is a standard benchmark in control literature and spiking control theories~\cite{vincentPositioning1989, slijkhuisClosedform2023, ahmadvandNeuromorphic2023}. Furthermore, the SMD as a linear system can be expanded to approximate more complex plants, as shown by its applications in muscle models for motor control and neuro-mechanical studies~\cite{byadarhalymodular2012, pazzagliaBalancing2025}. 
In our proof-of-concept, we make use of a simple target $\mathbf{z}$, which evolves during the simulation, incrementing twice and then stabilizing. In all of our simulations, the target state is always relative to the position of the mass, while its velocity can range freely. Although this is the setup we will work with, our algorithm allows to constrain the values of either the position or the velocity of the mass, by changing the entries in the cost matrix $\mathbf{C}$, as shown in Materials and Methods.
We will now illustrate the differences that emerge when controlling the system using a reactive versus a predictive paradigm, as described in Results (Fig.~\ref{fig3}). We consider both physically unconstrained control (where the control signal can affect both velocity and position), and physically constrained control (where the control signal can only affect the velocity).

\subsubsection*{Reactive control}

We first consider a physically unconstrained system with reactive control (Fig.~\ref{fig4}A). In this case, the system can be successfully made to track a target (Fig.~\ref{fig4}A top right). This is achieved by continuously spiking directly towards the target, which is always the better choice since the neurons do not consider future states of the plant.
In this example, the control is exerted by progressively kicking the mass in the same position as the evolving target, essentially `teleporting' it without affecting its velocity. Of course, this is only possible because we are considering a physically unconstrained system, where the mass is allowed to be moved instantly in any direction of the state space. Also note that because of this property, the system ends up in a physically impossible state with a non-zero velocity (green curve in upper right plot), despite remaining in place according to the position variable. 
We therefore ask if a reactive control method could work on physically constrained systems.
We employ reactive control to our physically constrained example, running the simulation of an SMD. In this context, we directly influence the velocity component of the state $\mathbf{x}$, while the position component evolves as a consequence. This reflects a key constraint of physical systems: the position cannot be instantaneously changed, as the mass cannot move from one position to another in no time.
As shown in Fig.~\ref{fig4}B, a reactive approach fails to control a physically constrained system. This is because influencing the velocity of the mass has no instant effects on its position. Therefore, the spikes can only take the velocity up or down in state space (see the red arrows in Fig.~\ref{fig3}A), increasing the distance from the target when compared to the non-spiking case (see brown dot Fig.~\ref{fig3}A). As a result, following our greedy spiking rule, the loss is never decreased by a spike, and the network remains silent. To appreciate these effects, the network should then be able to observe future states of the plant, accounting for the effect of the spiked velocity on the position of the mass.

\subsubsection*{Predictive control}
We introduce our predictive spiking algorithm to a non-constrained system. In this scenario, the controller is allowed to spike any dimension of the system, but it is also able to predict the plant's future states, and it produces spikes that improve the loss $f$ seconds from the current state. From Fig.~\ref{fig4}C, we can see how, when exerting the control, the position is sometimes directly moved towards the target (as in the reactive case), but optimizing the local losses often equates to kicking the system in its velocity, exploiting the intrinsic dynamics of the plant. Our predictive algorithm exploits the intrinsic dynamics of the system even in the absence of physical constraints, although in some instances, the mass is still instantaneously moved towards the target.
Finally, we apply the same predictive algorithm to the SMD simulation. In this scenario, the spiking network accounts for future states of the system, and is constrained to only spike the velocity either up or down, adjusting to the changing target position. The trajectory approaches the target and the loss $f$ seconds after a spike is minimized. If the position then departs from the target, the losses of the neurons will change their value, eventually causing the neuron to reach its threshold. At this point, the velocity will be either spiked up or down, causing the network to trace the target correctly.
As shown in Fig.~\ref{fig4}D, our predictive spiking rule introduces an element of anticipation, making the control strategy more robust in physically constrained dynamical systems, and allowing for the control of the SMD where a reactive control would fail.

\subsection*{The controller as a recurrent spiking neural network}
\begin{figure} [H]
\includegraphics[width=\columnwidth]{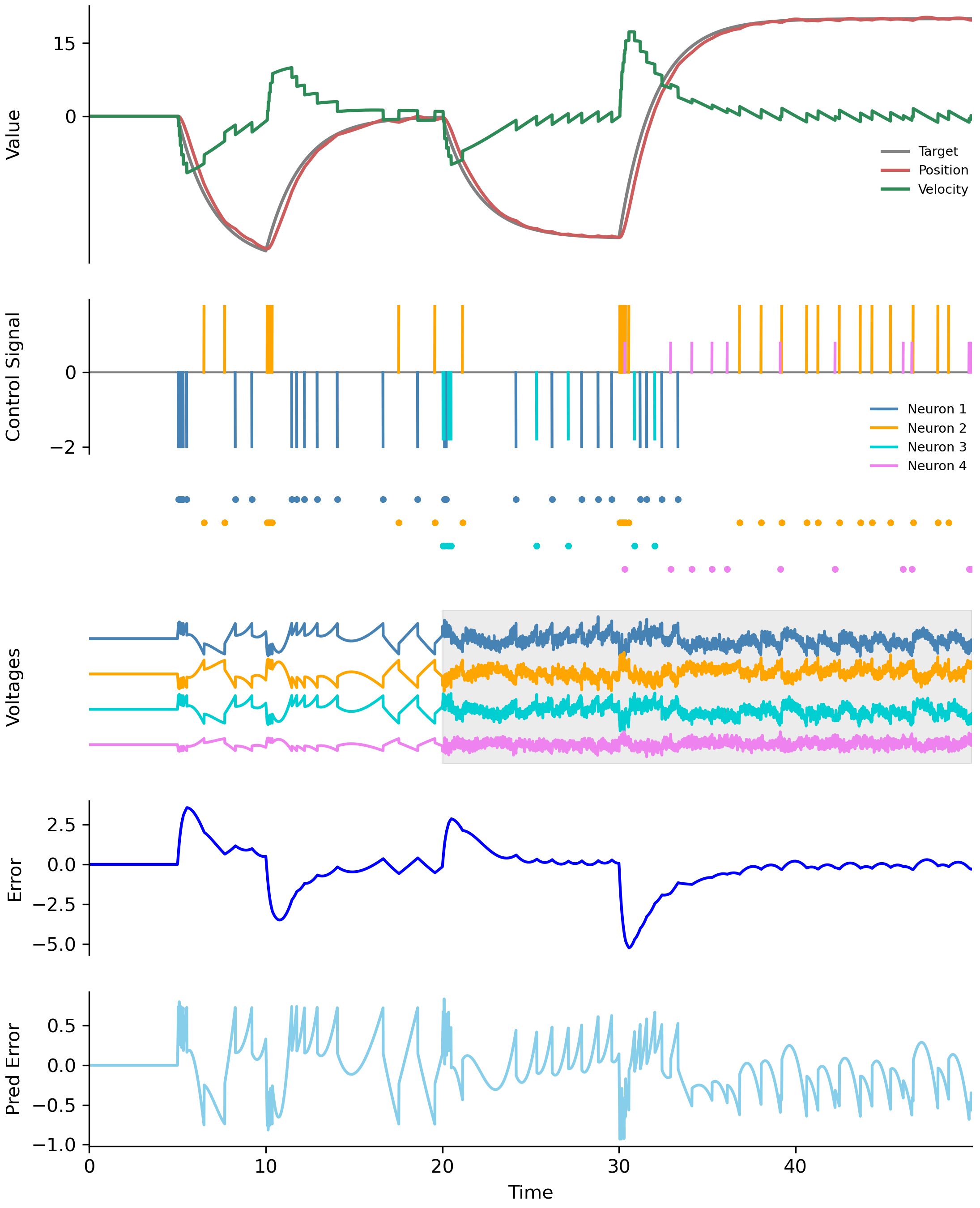}
\caption{ \textbf{Example activity of a network of four neurons.}
Each spike plotted as control signal is modeled as a Dirac delta function that gets mapped onto the plant's dynamics through the matrix $\mathbf{B}$, and instantaneously influences the velocity component of the system (see Results). Spike moments and the relative voltage traces are shown, as well as the resulting errors. The error trace (deep blue line) describes the current distance between the position and a target state, while the predictive error (light blue line) measures the distance between the target and a predicted position, $f$ seconds into the future. At time $T=20$ seconds into the simulation, we inject noise (with standard deviation $\sigma=0.08$) in the voltages of the neurons. The network remains robust to noisy spiking and the control performance is not affected.} 
\label{fig5}
\end{figure}

We now focus on the example activity of a predictive control of an SMD with a network of four neurons, as shown in Fig.~\ref{fig5}. A visual representation of the structure of this network is shown in Fig.~\ref{fig2}. In this example, we want to show the network's performance, spikes, voltage traces, and errors. To bring all four neurons to spike, we choose a slightly more complex target $\mathbf{z}$, which spans positive and negative values. The control task and the network are set up in the same way as the previous SMD example (Fig.~\ref{fig4}D).

In each timestep of the simulation, the neurons compute spiking and non-spiking losses, accounting for the future state of the system in $f$ seconds, and its distance towards the target. 
The distance to the target considered in these losses is represented as the predictive error. When its value is large enough, the neuron's threshold is reached, and a spike is produced to minimize it. In turn, this minimizes the actual error, allowing the position of the mass to approach the target position. 
The sudden changes in value of the predictive error are caused by the voltage traces of the neurons reaching their threshold, and thus temporally coincide with the spike events. 
Each spike is a direct control input. It is scaled by the matrix $\mathbf{B}$ by a different amount for each neuron, and mapped as a positive or negative contribution to the velocity in $\mathbf{x}$. Between spike moments, the system evolves as an open loop system, without any control signal in input, following its intrinsic dynamics. The control loop is closed each time a spike is produced and mapped onto the velocity. Since each spike is modeled as a Dirac delta function, the corresponding change in velocity is instantaneous and the control naturally exploits the intrinsic dynamics of the plant to minimize the predictive error. At time $T=20$ seconds, we inject noise in the voltage of each neuron, which can cause them to reach their threshold independently from the computed loss. The local nature of the spiking rule helps with preventing this noisy spiking to affect the resulting control performance.

\subsection*{Varying meta-parameters}

The control network we employ operates by expressing its neural parameters with elements of the downstream plant, which serves as a proof-of-concept for a spiking network that can directly compute control in a linear dynamical system. 
However, at this stage, other features of the network are still imposed as meta-parameters to adjust the control to our SMD example. For a detailed list of all the parameter setup used in this work, see Materials and Methods.

\paragraph*{Number of neurons ($N$)}
The number of neurons in the network heavily influences the control of the plant. For instance, when operating on a system with very simple oscillatory dynamics as our SMD, employing more neurons can make the control more robust to different settings, but it raises the need to manage the spiking activity among neurons, especially since they do not have access to network-level information. This can be achieved by imposing asynchronous firing or introducing an adaptive cost on spike, as mentioned in Results and Materials and Methods. While for our activity example in Fig.~\ref{fig5} we showed the results for four neurons, in all other demonstrations we kept the network as simple as possible and employed only two neurons: spiking the velocity of the plant either up or down by the same amount. Indeed, for systems with more complex dynamics and targets, exerting effective control requires larger networks, as in the coupled oscillator case of Fig.~\ref{fig8}, where $N=500$.

\paragraph*{Kick strength ($\mathbf{Bs}$)} 
The spikes of our SNN are Dirac delta functions that operate as control signals on the dynamics of the plant. These signals are mapped onto the downstream system via the matrix $\mathbf{B}$, which can arbitrarily scale their value, resulting in smaller or larger kicks to the velocity of the system. This is the case for Fig.~\ref{fig5} and Fig.~\ref{fig8}, where each neuron kicks the velocity differently. In all the other examples, since we refer to the same SMD system with the same dynamics, we decide to keep $\mathbf{B}$ fixed, only allowing the system's velocity to be kicked up or down with the same strength. This makes our proof-of-concept as clear as possible. Of course, the value of $\mathbf{B}$ holds great relevance when varying the control objective, as it can let the same network control significantly different plants, as well as exert control on the same plant with different levels of accuracy. For this parameter exploration, since we only tackle the same SMD example, we keep a fixed $\mathbf{B}$ value that produces kicks of a compatible magnitude with the downstream plant.

\begin{figure} [H]
\centering
\includegraphics[width=0.8\textwidth]{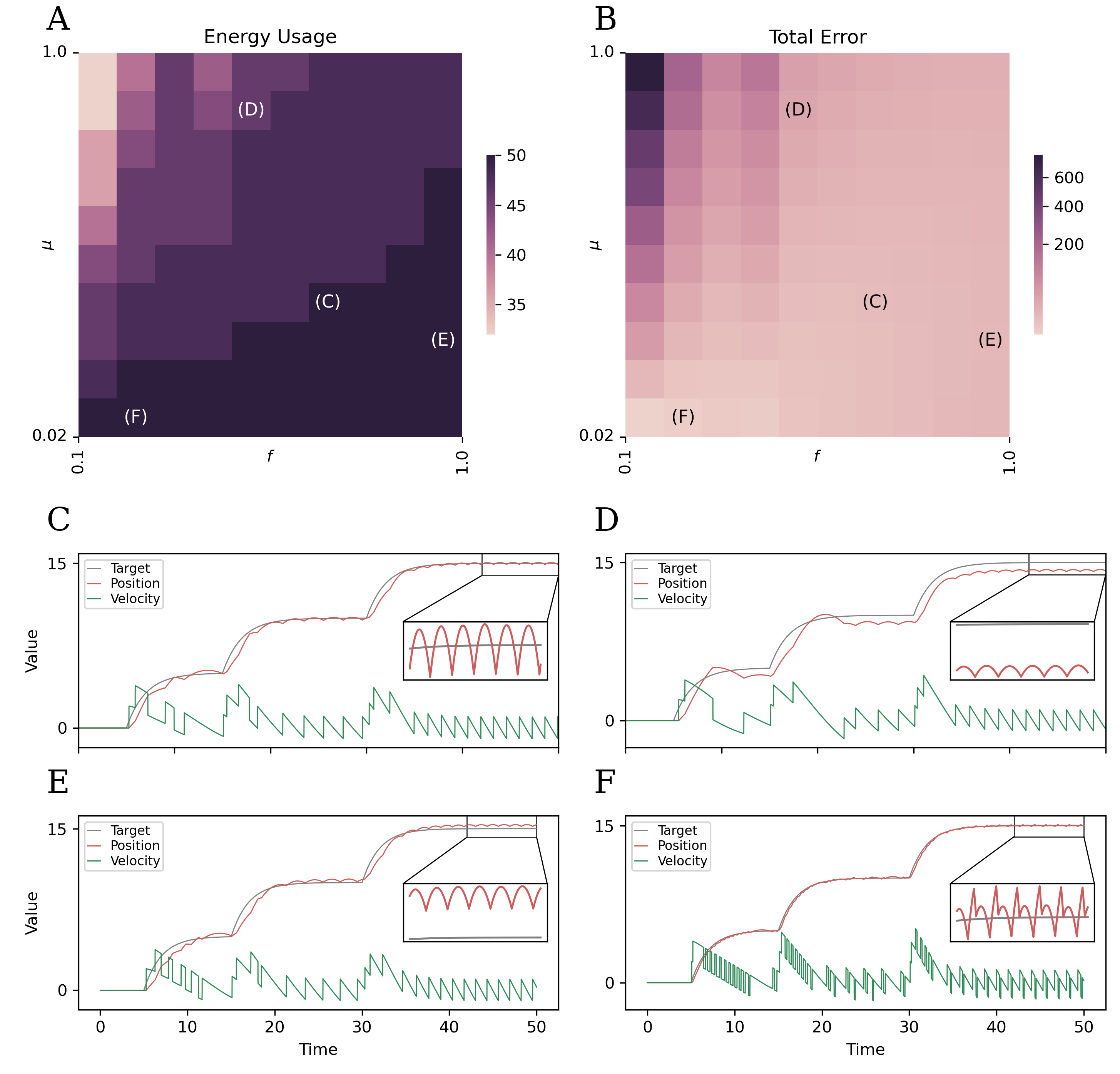}
\caption{\textbf{Parameter sweeps and control performance.}
Changing two meta-parameter ($\mu$ and $f$) in simulating SMD with spiking control.
A: Energy Usage, calculated as effected acceleration onto the system in each simulation ($\mathbf{Bs}$), for various combinations of $\mu$ and $f$ (linear scale). As described in the Appendix, the measure of acceleration is chosen for simplicity of interpretation in our framework. However, measures of work done on the plant can also be adopted to describe the energy input in the system.
B: Total Error, calculated as cumulative distance from the state $\mathbf{x}$ to the target $\mathbf{z}$ in each simulation, for different combinations of $\mu$ and $f$ (logarithmic scale).
C-F: Different control regimes for different combinations of $\mu$ and $f$. C: efficient control regime. D: under-control regime. E:over-control regime. F: inefficient control regime.}
\label{fig6}
\end{figure}

\paragraph*{Global threshold scaling ($\mu$)}
When deriving the network, we can impose a scaling term $\mu$ on the loss, to introduce a cost for spiking (Eq.~\ref{eq:control_loss}). This term influences the neuron's threshold value (see Materials and Methods). Scaling the neuron's threshold gives interesting results when controlling the plant (Fig.~\ref{fig6}), since it affects the number of spikes emitted as well as the accuracy in tracing the target position. Parameter $\alpha$ has a similar effect on the resulting control, since it scales the threshold based on the neuron's past activity.

\paragraph*{Predictive window ($f$)}
When computing the spiking decision, each neuron compares spiking and non-spiking losses. As mentioned, these are predictive losses: they take into account the effect of a spike on the system $f$ seconds into the future (see Results). The width of this future time window is crucial for the control results (Fig.~\ref{fig6}). Looking further into the future allows neurons to better assess whether the trajectory will approach or depart from the target, often preventing unnecessary spikes. On the other hand, 
if the window $f$ is very large, the predicted state used in the loss comparison might approach (or depart from) the target way before the actual state of the plant does. This influences the accuracy with which the network represents the effect of its own control on the system's dynamics, as we will explain below.

\end{multicols}

\begin{multicols}{2}

\subsubsection*{Control regimes}
Different combinations of these meta-parameters can give different results in the control of the plant. Quantifying the optimality of the control strategy in comparison to other paradigms (e.g. standard LQR, or filtered-spikes) is very difficult. As described in the Discussion, each neuron in our network follows a greedy spiking rule given the control set, but like in other approaches, this does not necessarily produce the best overall controller. It is of interest to vary some control parameters, considering the cases where the changes in target value are traced by the system's state (and therefore the SMD is successfully controlled). Within these cases, we identified different regimes of control for our network. In these explorations we decided to maintain fixed the number of neurons ($N=2$) and the kick strength $\mathbf{B}$.
We therefore vary the future time window $f$, influencing the predicted state based on which the network computes the spiking decision, and $\mu$, the global scaling factor for threshold.
As shown in Fig.~\ref{fig6}A, networks with a lower $\mu$ tend to spike more, using more energy, as defined by the effected acceleration onto the system (for this system, this equates to the sum of the total spike count of the simulation, with each spike scaled by the corresponding value in the matrix $\mathbf{B}$). Given the same $f$, these networks also tend to have a lower error compared to high $\mu$ networks (Fig.~\ref{fig6}B), since a lower threshold allows to adjust more often to the target position. In general, it emerges from Fig.~\ref{fig6}B how a relatively wide range of parameters allows for the control of the plant, giving comparable error measures. Only more extreme parameters combinations (a high threshold scaling, especially with small $f$) result in the network not tracing the target, yielding significantly higher cumulative error.

\paragraph*{Efficient and inefficient control}
Within the boundaries of these trade-offs, we identified four regimes (Fig.~\ref{fig6}C-F). With an adequate future time window $f$ and relatively low $\mu$, the network can operate in an efficient regime (Fig.~\ref{fig6}C, where the target position is followed precisely, while keeping a relatively low spike count. If the spike count is lowered any further from this point, the control input becomes too sparse, and the SMD mass does not closely orbit the target position. On the contrary, decreasing the value of both these meta-parameters yields the inefficient regime, where the target is still traced with comparable precision, but the spike count, and as a consequence the energy input, is considerably higher (Fig.~\ref{fig6}F).

\paragraph*{Under- and over-control}
We then have the case where the network is allowed to look far enough in future states of the system, but $\mu$ is also very high, making it difficult for the neurons to spike enough, and to trace the target accurately (Fig.~\ref{fig6}D). In this regime of under-control, the position approximates the target but never manages to reach it. 
Finally, in the last case, the neurons are relatively free to spike as $\mu$ is low, and they are bound to look very far into future states of the system (Fig.~\ref{fig6}E). In this regime of over-control, the neurons fire based on a predicted position that is far into the future (large $f$) and thus differs significantly from the actual one. For instance, while the current state of the system might have not reached the target yet, the predicted state is already well past it, which causes the neurons to fire earlier than needed. This over adjustment makes the position hover slightly past the actual target. 
\end{multicols}

\begin{figure} [H]
\centering
\includegraphics[width=0.87\textwidth]{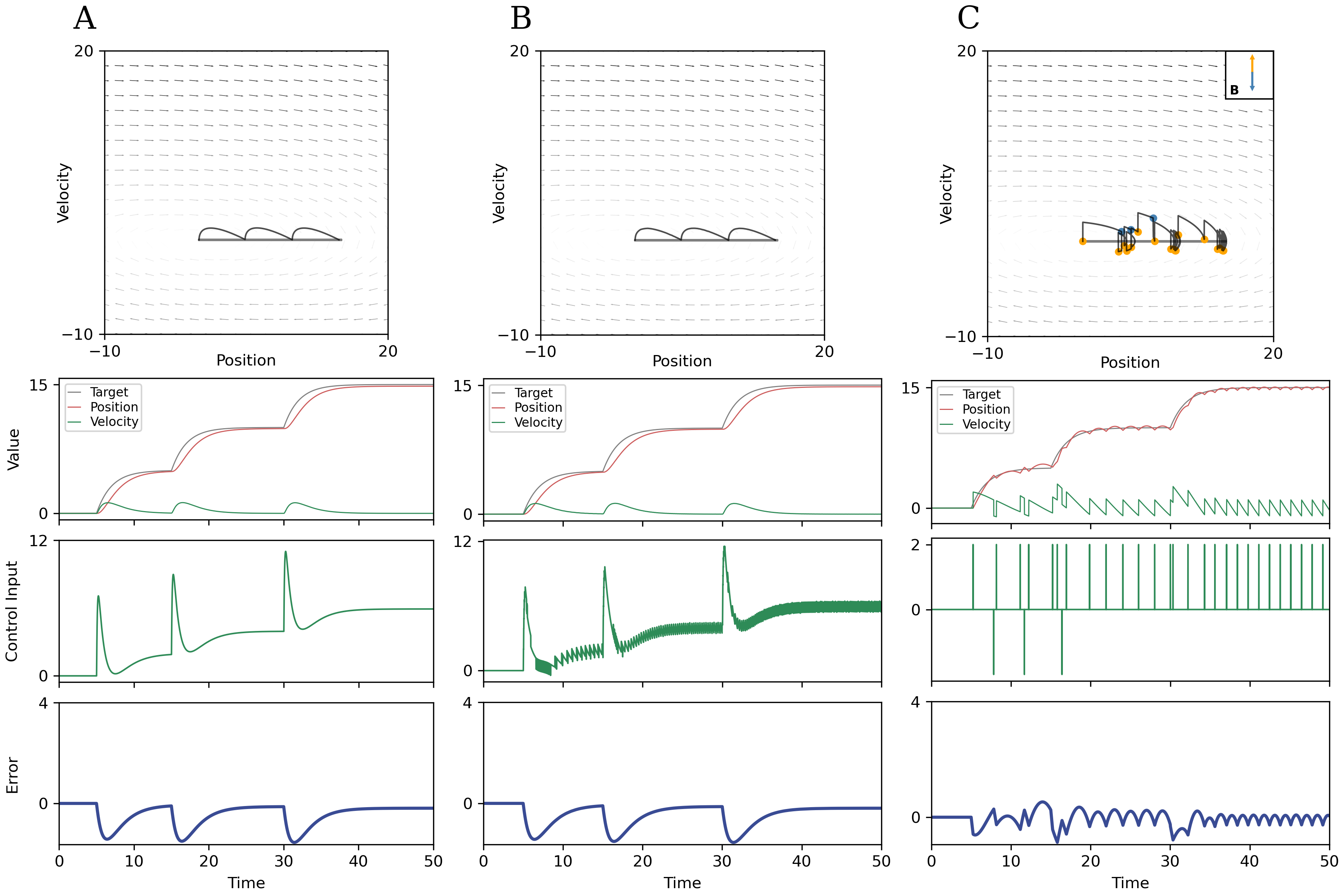}
\caption{\textbf{Control example of an SMD with different approaches.}
Simulation of the control of an SMD with continuous control (LQR), filtered-spikes control, and predictive spiking control:
A: LQR control example: the controller continuously adjusts the position to the target by modulating the velocity. This is shown in the phase plane (first row) and in the activity plot (second row). Each time the target value changes, the position slowly adjusts to it, stabilizing when the target is reached. The control signal (third row) is larger when the target changes value, and the velocity is modulated to adapt to it. During these moments, the mass is the furthest away from the target, which is reflected in the error profile (bottom row).
B: The filtered spikes control is approximating the LQR signal. Therefore, the trajectory in space, the activity over time and consequently the error (first, second, and last rows) have the same profile as in panel (A). The approximation is seen in the control input (third row). Since the system is represented through spikes, the network keeps spiking to trace the LQR signal. 
C: In our predictive spiking algorithm, spikes directly influence the velocity of the system. This is shown in the upwards and downwards kicks in state space and in the activity plot (first and second row). On the one hand, this helps the controller adjust faster to changes in target values. On the other hand, when the target is reached, the controller can only orbit its position by spiking. This fundamental difference is also highlighted in the error profile (bottom row). In the control input plot (third row) we can see how the spike count is drastically lower in spike control than in the filtered-spikes one.}
\label{fig7}
\end{figure}

\begin{multicols}{2}

\subsection*{Comparing control approaches}
Lastly, we want to showcase the differences in the resulting control for three approaches: the continuous LQR control, the filtered-spikes control, and our predictive spiking control. As for previous examples, we consider a physically constrained simulation of an SMD, with a progressively changing target. In Fig.~\ref{fig7} we directly compare the same measures for all three methods. As shown in the error trace (bottom row), the filtered-spikes algorithm (Fig.~\ref{fig7}B) approximates the control strategy of the continuous LQR (Fig.~\ref{fig7}A), as described in Materials and Methods. Spiking control (Fig.~\ref{fig7}C) has similar increases and decreases of the error (contingent to the changes in the value of the target), although its profile is very different, as no optimal control signal is being traced. Furthermore, the non-continuous nature of its control inputs only allows the position to orbit the target with a certain accuracy, without ever stationing on the exact target value. Analogous differences are highlighted when showing the activity over time of these networks (Fig.~\ref{fig7}, second row). 
In the continuous and filtered cases, the velocity is not spiked, and the position slowly adjusts to the changes in target, stabilizing once the target is reached. On the contrary, spiking control is progressively adjusting the velocity up or down, tracing the target as soon as it changes value. Once the target is reached, the controller orbits the target position by spiking regularly until the end of the simulation. 
Other fundamental differences are highlighted in the control input (Fig.~\ref{fig7}, third row), and in the consequent trajectory in state space (Fig.~\ref{fig7}, first row). As shown, the filtered spikes controller is able to smoothly control the state of the system by approximating the control input of the LQR. In this sense, the spikes are only representing the system's state, but the control signal is still continuous, as it is inputted in each timestep of the simulation. This approximation requires the network to spike much more than our spiking control algorithm, which exclusively inputs a kick when it needs to adjust the position to the target, and is not approximating any optimal control solution. Finally, on the comparison between filtered and spiking control, the role of a filtering element in the paradigm depends on the modeling choice we operate and the hypothesis we consider for the specific research question. In the filtered spikes approach, the filter is present within the network itself, so that the spikes represent but do not compute the control signal directly. In our spiking algorithm, the filtering element is placed outside of the network and into the plant: here, the matrix $\mathbf{B}$ directly maps spike events onto the system state, which makes the control inputs impulsive rather than continuous. Indeed the choice of positioning and defining a certain filtering element in different component of the control paradigm is partly artificial, but it does determine the nature of the control input and the interpretation of each element of the paradigm itself. As described in the Appendix, there is no universal energy advantage of our spiking algorithm from a control-theoretic perspective when compared to filtered spikes or continuous approaches. Rather, the strength of our contribution lies primarily in the formulation of the spiking control scheme, which guarantees sparsity of the resulting spiking activity, interpretability of the model's components, and adaptability to spike-dependent control applications (for instance, when using neuromorphic hardware).

\subsection*{Controlling coupled oscillators}
\begin{figure} [H]
\includegraphics[width=\textwidth]{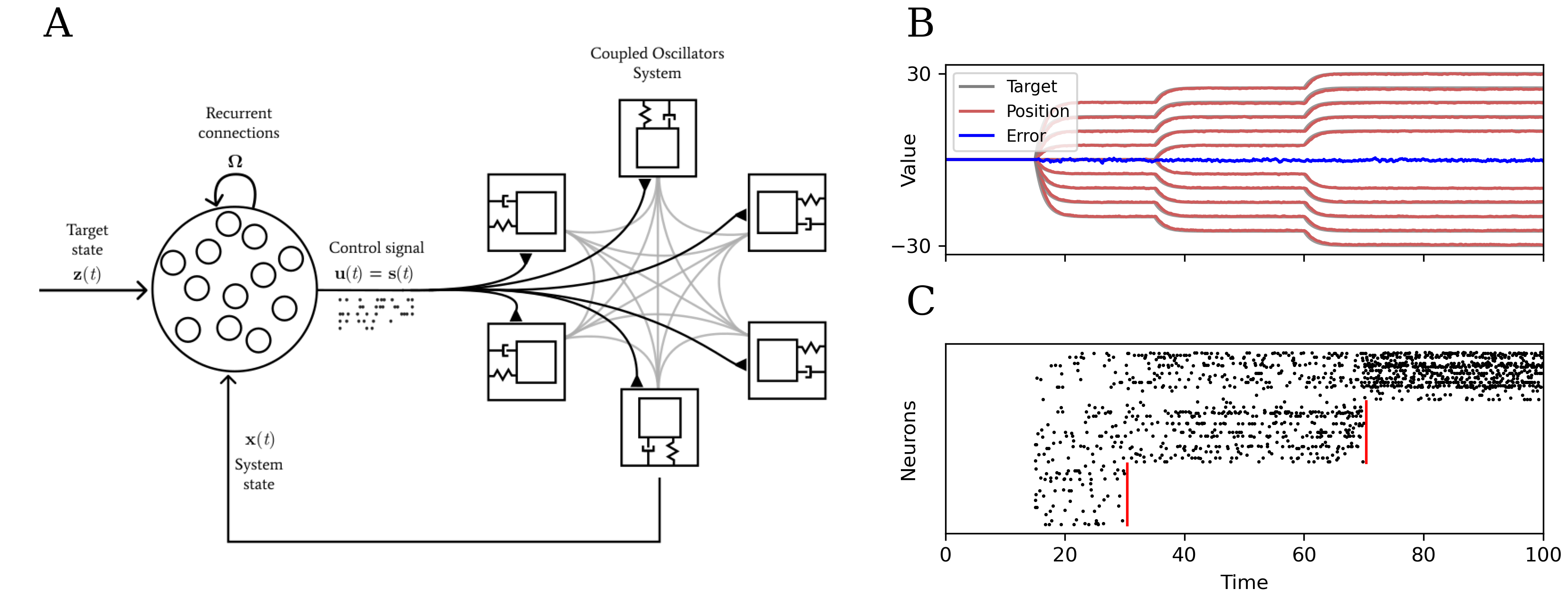}
\caption{\textbf{Control simulation of a coupled oscillator system.}
A: Scheme of the feedback spiking control for a downstream coupled-oscillator system. The current state of the system $\mathbf{x}\mathit{(t)}$ is given as input to the network, along with target state values $\mathbf{z}\mathit{(t)}$. The SNN outputs spikes $\mathbf{s}\mathit{(t)}$. These represent the control signal $\mathbf{u}\mathit{(t)}$, computed as a function of both inputs. As for the other networks, $\mathbf{u}\mathit{(t)}$ is then mapped onto the dynamics of the downstream system via the $\mathbf{B}$ matrix. This time, the downstream system is composed of $M$ masses connected to each other, such that the acceleration of each mass influences the position of the next one. The plant dynamics are still linear, but they are significantly up-scaled in dimensionality when compared to the single SMD examples.
B: Activity for the control of a system with $M=10$ coupled masses. Each mass has a different target position over time $\mathbf{z}\mathit{(t)}$. For this example, the SNN makes use of a randomly initialized $\mathbf{B}$ matrix to map the spikes onto the high-dimensional system and guide the position of each mass to its designated target. Note: for clarity, here we do not show the spiked velocity over time like in previous figures. In blue we show the cumulative error of all neurons over time.
C: Spiking activity of the controlling network. We demonstrate the robustness of the control by progressively silencing portions of the $N=500$ neurons during the simulation. The vertical red lines indicate the silencing of neurons in certain timesteps. The network successfully redistributes the spiking activity among the non-silenced neurons and as a result, the control performance is not affected (as shown in panel B).}
\label{fig8}
\end{figure}

\begin{figure*} [!ht]
\centering
\includegraphics[width=0.8\textwidth]{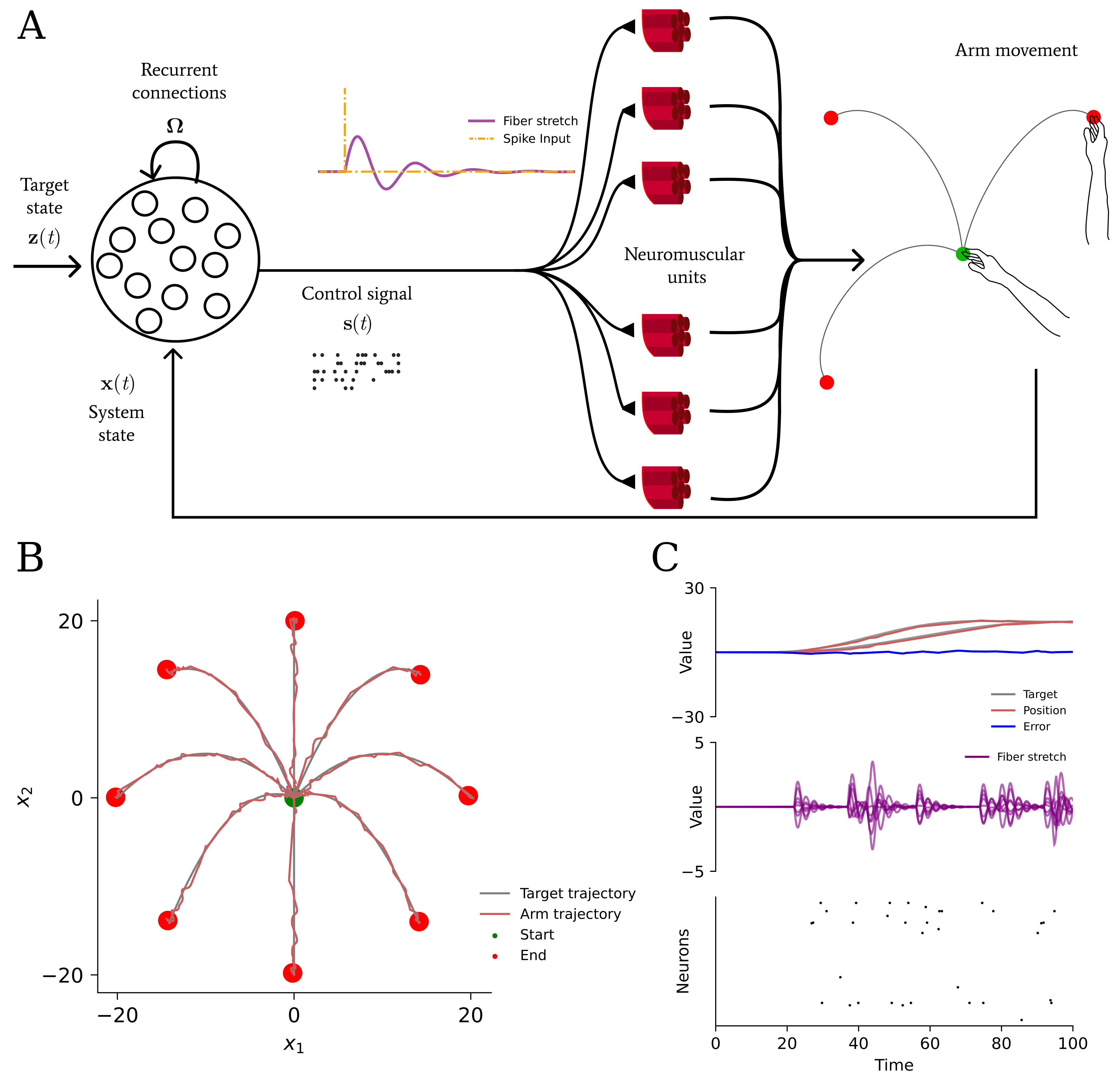}
\caption{\textbf{Arm reaching simulation} A: Scheme of the
feedback spiking control for a single point arm reaching task. The current state of the system $\mathbf{x}(t)$ includes the states of the neuromuscular units and of the point arm. It is given as input to the network, along with target state values $\mathbf{z}(t)$ which are only relative to the arm position (the state of each u
nit is not constrained by the target). The SNN outputs spikes $\mathbf{s}(t)$, are then mapped through $\mathbf{B}$ onto the dynamics of the fiber units. The spikes affect the velocity component of each SMD, causing an oscillatory displacement. In these examples, the $N=500$ neurons are randomly assigned to eight uncoupled SMDs. The state of such units are then linked to a 2D point mass. B: Examples of eight separate reaching tasks where the network controls the fiber units to guide the point arm through eight different target trajectories. C: Activity measures for one of the eight tracing tasks. In the top panel, the two coordinates $x_1$ and $x_2$ of the point arm correctly trace the target trajectory, as the error is kept around zero. In the middle panel, the eight SMDs oscillate according to the spike input, and influence the arm movement. In the bottom panel, little spike activity is needed to generate the necessary oscillations in the unit and cause the point arm to move in space.} 
\label{fig9}
\end{figure*}

We now scale up the same algorithm described in previous sections to control a higher dimensional linear system. This time, the plant we chose is a coupled oscillator system, where $M$ number of SMDs are connected in series, such that the position of one mass affects the acceleration of the ones connected. The network has to control multiple SMDs, while taking into account the coupled effects between them. For the example shown in Fig.~\ref{fig8}, we consider the control of $M = 10$ coupled masses using a network of $N = 500$ neurons. As shown, the control algorithm scale naturally to this complex linear system: even though the masses are coupled with each other, our local spiking rule is able to coordinate the spikes as to allow all masses to reach their respective target (Fig.~\ref{fig8}B). Neurons are assigned an element in the matrix $\mathbf{B}$ (now scaled up to the correct dimensionality based on the number of masses $M$) with a random normalized value. The spike of each neuron then maps either as a positive or negative kick on the velocity of a mass by the corresponding amount. This results in some neurons having a higher kick strength than others, but as before, only the neurons that allow the system to decrease the distance from the target in $f$ second will fire. During the simulation we observe how spiking activity is sparse and distributed among neurons (Fig.~\ref{fig8}C). We then silence random neurons to test the robustness of the system. After each silencing event, the activity is distributed among the remaining active neurons and there is no effect on the control performance. This shows how our control algorithm scales to complex linear system and is robust to cell loss: given a large enough network, no specific neuron is determining the outcome in the control performance. The robustness observed in this system is a direct consequence of the greedy spiking rule and the redundancy inherent in the SCN framework, described in \cite{boerlinPredictive2013}. In this architecture, the membrane potential of each neuron tracks a projection of the global prediction error (distance to the target). If a specific neuron is silenced, the prediction error it would have corrected continues to accumulate in the membrane potentials of the remaining active neurons. Thus, neurons with a similar alignment (in our case, the neurons whose spikes are closely mapped in state space) will fire, redistributing the work across the remaining neurons in the population.
Another example of spiking control over more complex linear system can be found in the Results, where we employ our algorithm to control a series of muscle fiber unit models to move a point arm in space (Fig.~\ref{fig9}).

\end{multicols}

\begin{multicols}{2}

\subsection*{Arms reaching task simulation}
We now use our paradigm to propose a bio-inspired example of motor control.
In this example, the spiking network aims at controlling a point arm model through oscillating SMD elements that represent simplified fiber units. The use of controlled oscillators as models of muscle fiber displacement is present in the literature ~\cite{byadarhalymodular2012, bernikernormative2019, pazzagliaBalancing2025}. The overall dynamical system is described by a single linear state equation, whose state vector $\mathbf{x}(t)$ contains both the kinematic variables of the arm (its position and velocity in time) and the internal mechanical states of the SMDs. Although all the state variables transformations are included in the same system matrix $\mathbf{A}$, they play fundamentally different physical roles. The arm states correspond to observable coordinates of a point-arm extremity moving in two dimensions. In contrast, the SMD states represent internal muscle-like fiber extensions (fiber stretches) relative to their equilibrium lengths. These displacements are not spatial coordinates in the external workspace; rather, they quantify mechanical deformation within each fiber unit. Through standard SMD dynamics, a fiber extension and its velocity generate force, which in turn drives the arm’s motion. The spikes of the network directly control the velocity with which the fiber is extended (or contracted). The SMD dynamics then evolve continuously according to their intrinsic spring stiffness, damping, and inertial properties. The collective state of these units is then linked to the position and velocity of the point arm in space (Fig.~\ref{fig9}A). Importantly, the control objective (and consequently the loss) is defined exclusively in terms of the arm’s target trajectory. The target $\mathbf{z}(t)$ specifies desired time-varying coordinates in two-dimensional space, and no explicit target is assigned to the SMD states. Although the full system state includes both arm and fiber units variables, the cost function penalizes only deviations of the arm position from the desired trajectory, leaving the internal unit states unconstrained except through their mechanical coupling.
In the example provided in Fig.~\ref{fig9}B, eight units oscillate in response to spike control signals and consequently affect the position of a single point in 2D space to trace a certain target. Starting from a point in the origin (green dot) we set eight different target trajectories to be traced consecutively and their respective end points (red dots). As shown, the network successfully controls the state of the fiber units to trace the target movements. 
In Fig.~\ref{fig9}C, we show some measures of activity for one of the example trajectories. We illustrate how both components of the trajectory ($x_1$ and $x_2$) are correctly traced, and the error is minimized. This is caused by the fiber displacement (stretch) of the units over time. Each spike mapped onto these units is reflected in a dampened oscillatory behavior. As a result, as shown in the bottom panel, very sparse spiking behavior is sufficient to generate the correct SMD responses that drive the point arm to the target.

\section*{Discussion}

\subsection*{Proposed spiking control framework}
As described in Introduction, biological neurons use spikes for signaling and control of downstream system.
Control of dynamical systems has been widely studied and applied in different contexts, including artificial neural networks ~\cite{boerlinPredictive2013, deneveEfficient2016, huangDynamical2018, slijkhuisClosedform2023, mooreneuron2024}, but generally doesn't consider spikes directly as control signals. 
In this work we remedied this by showcasing how control can be computed directly through the use of single spike events. This allowed us to obtain a model that uses spikes to actively kick the dynamics of the downstream plant, and exploits its intrinsic dynamics to reach a certain target.
The emitted spikes are instantaneous events (modeled as Dirac delta functions, see Results). The spike generation mechanisms in the network and the mappings to the system need to be adjusted to the various constraints the plant might have in its dynamics (as explained in Results). In order for this control paradigm to work on physically constrained systems, we developed a predictive algorithm, that takes into consideration the effect of our control inputs onto the system within a certain future window, $f$. 
Our spiking rule is fully local and is computed for each neuron in every timestep of the simulation. Therefore, the target can be reached without a network-level objective, and without any explicit learning paradigm. Of course, introducing learning could still be a very useful addition to adjust all the meta-parameters of the network when adapting the control to different plants, or to tackle plants with unknown dynamics.
From this simple spiking rule, we obtain, without imposing it, a recurrent network of IF neurons that exerts control by only inputting a spike if it improves the defined loss. This rule does not guarantee the optimality of the control as a whole (as explained in Discussion), but allows for the control to work without the need for network-level information. Lastly, we scaled up this framework to control a coupled oscillator system, with various connected masses (each with its different target) and a larger network. We showcase that not only our control algorithm is easily scalable to complex linear systems and higher-dimensional networks, but it is also robust to neuron loss.

\subsection*{Assumptions and constraints}
\paragraph*{Greedy spiking rule}
An important clarification about our proposed algorithm is that the spiking rule by which the network produces control signals does not guarantee any optimality on the controller as a whole. Each timestep in the simulation, the neurons operate the spike decision by simply trying to decrease the distance towards a target. That is why the neurons produce control signals (spikes) in a ``greedy" way, as defined in the work of  \cite{boerlinPredictive2013}. If spiking improves the loss, the neuron always spikes, as that is the only information that it has available. No other network-level information is used in the process.
In other words, computing the greedy spike decision for each neuron in the current timestep (albeit considering the effect of the spike in $f$ seconds), does not equate to finding the best set of spikes over a whole future window, let alone over the whole simulation. 
As a result, our approach does not guarantee that the control actuated by the network is necessarily the optimal control solution for reaching a certain target. Furthermore, as described in the Results, many of the meta-parameters in the network are manually set rather than learned or optimized for the control task at hand. In most of our examples, we impose the meta-parameters that best fit the scale of the system to control, to showcase our proof of concept with clarity.

\paragraph*{Features of the control scheme} 
From the described spiking rule, we obtain a recurrent LIF network able to exert control over a linear dynamical system. Although such spiking rule is based on a loss evaluated over a time window of duration $f$, the control problem considered in this work is of infinite-horizon in nature. That is to say, there is no terminal time or terminal cost in the control problem, and control actions are generated progressively as the system evolves, without reference to a final target state. Therefore, the parameter $f$ only specifies a fixed-time prediction window over which the loss is evaluated. The control decision at each time does not involve any optimization of future control actions.\\
Consequently, this work does not directly include an account for how the cumulative effect of spikes allows for the reduction of control error. However, the stability analysis provided by \cite{eilersStability2025}, traces analogies between our discrete impulses control method and an analogue continuous control, and in doing so, proves the control stability of our paradigm. The core of the work of \cite{eilersStability2025} is the introduction of an auxiliary variable $x_c(t)$, defined as the sum of the plant state $x(t)$ and a linear combination of the internal neuronal variables $z(t)$ (analogue to our voltages $\mathbf{V}(t)$). The auxiliary variable can be seen as a continuous counterpart of our impulsive control, accounting for hypothetical impulses from the neurons, rescaled by how close their voltage is to its threshold. Applying these hypothetical impulses to the system state amounts to a jump from $x(t)$ to $x_c(t)$. In the section "Extension to connected neuronal units" of their work, \cite{eilersStability2025} use Lemma 12 to prove that the internal neuronal variables $z(t)$ of the network remain uniformly bounded from above and below. Because the neuronal variables are bounded, the ``perturbation" they introduce to the continuous auxiliary variable $x_c(t)$ is also bounded. Since the neuronal variables are encoding for the distance between the state and the target, that error (while not reaching zero due to the kicks strenght $\mathbf{B}$), will converge to and stay within a small region around the target. The spiking system is then guaranteed to be practically stable in the control of the plant.

\paragraph*{Simplicity of use case}
Following a similar reasoning, we purposefully chose a simple dynamical system, which is linear (fully observable) and two-dimensional. We also chose a simple target that progressively changes position and then stabilizes. This simplicity allowed us to showcase our results without any additional complication, demonstrating how this framework easily scales to any linear dynamical system. However, the limits of our paradigm still need to be tested, by including adaptive (or learned) meta-parameters, and expanding our control algorithm to operate on non-linear systems (see Discussion). 
Important to note that even in an optimal controller, these meta-parameters would need to be adjusted based on the control problem at hand.
The differences in the control approach of our network is highlighted in Fig.~\ref{fig7} and explained in Results, where we control the same SMD system with a standard LQR (optimal controller), a filtered spiking algorithm, and our spiking controller. As shown, the filtered spiking controller (Fig.~\ref{fig7}B) approximates the control signal of an LQR (Fig.~\ref{fig7}A), making the resulting trajectories in state space equivalent, while our spiking algorithm (Fig.~\ref{fig7}C) differs significantly from the optimal trajectory in state space. However, by adjusting the meta-parameter correctly, the network still traces the target with good accuracy (as shown in the error profile), while producing significantly less spikes than the filtered spiking controller.

\subsection*{Considerations on spiking control}
\paragraph*{Relation to prior spike-based control models}
Previous works underlined the importance of each spike event in the computation of neural networks~\cite{boerlinPredictive2013, schwemmerConstructing2015, brendelLearning2020}. Such spike-based computational models highlight the relevance of the temporal precision of spike events, where precise spike timing is determined by each neuron’s contribution to a desired output. These models exhibit features similar to cortical networks, including irregular spiking and E/I balance~\cite{schwemmerConstructing2015}. 
Many of these models are developed from SCN theory, which gives them a strong overlap with our framework but also introduces important distinctions (as described in Materials and Methods). In SCNs, as in our approach, spike activity is used to generate a signal that controls a dynamical system.  However, SCNs use this signal to control their own network read-out (thereby influencing the dynamics of their internal state). This read-out can be quite complex, and include synaptic kernels, requiring a form of predictive spiking~\cite{schwemmerConstructing2015, zeldenrustEfficient2021} similar to the one employed in our paradigm. 
In short, despite SCN-related works share common aspects with out approach, they do not explicitly hypothesize that the computational role of spikes could be controlling a different downstream system. In contrast, our work generalizes to the control any downstream linear system, rather than focusing solely on modulating the network’s own internal state. Moreover, when SCN principles are employed to solve realistic control tasks~\cite{huangDynamical2018, slijkhuisClosedform2023, mooreneuron2024}, the emphasis on a temporal, single-spike perspective is lost, favoring a rate-based or filtered-spike approach, which glosses over the role of each spike in the network's computation. 
\paragraph*{Future outlook}
In neuroscience, single spike events for control could have a huge significance when modeling downstream actuator systems. Spike-based control seems to be a promising modeling angle when explaining the control of downstream muscle actuators~\cite{bernikernormative2019}, that ultimately allow movement to occur. This paradigm could be expanded to include models of body orientation in space as well as the control of more detailed models of limbs ~\cite{eliasmithunified2005}. Spiking control is also compatible with existing dynamical system approaches to modeling complex inter-network phenomena like epilepsy \cite{qinAnalytical2024}. Further down the line, spiking control could be a useful approach to interpret how the spikes influence the dynamics of the network that produce it. This has been theorized by previous works in SCNs~\cite{boerlinPredictive2013, deneveEfficient2016}, suggesting that the computational role of spikes in the network is to control its dynamics, and many physiological features of biological networks are a consequence of this necessity. In this case, the downstream system is the network itself (or a second downstream network), which means that our paradigm could serve as a tool to explain potential roles of the spike code in all bio-plausible networks. As mentioned above, various works emerged from this initial SCN perspective to investigate the biophysical and computational properties of these networks, and their potential values for interpreting detailed neural models~\cite{schwemmerConstructing2015, zeldenrustEfficient2021}. Not only these works stem from the common root of SCN-based modeling, but they also incorporate elements that we employ in our algorithm, such as different forms of predictive spiking. Revisiting these topics with our control-focused perspective could give interesting insights, for instance, on the computational role of E/I balance~\cite{podlaskiApproximating2024} or provide 
fresh perspectives on various properties of SNNs~\cite{mancooUnderstanding2020}. Following a similar path, we could study the effect of bio-plausible additions in our framework, like noise or network perturbations, analyzing for instance the robustness of our system~\cite{calaimgeometry2022}. 

Expanding our proof-of-concept to more complex nonlinear systems could be relevant to investigate the role of spikes in controlling realistic downstream actuators. This could include detailed models of muscle fibers for the control of movement~\cite{bernikernormative2019}, or models of complex downstream networks. Furthermore, it could have use cases for neuromorphic control applications where a better exploration of the full state space of the plant is needed. Tackling nonlinearity might require further refinements on the model by employing local linearization techniques~\cite{yesildirekFeedback1995, baratchartlocal2009} or a simultaneous prediction of multiple future states ~\cite{nardinNonlinear2021}.
In the effort to apply this algorithm to more realistic use cases, another required step would be to find a way to adjust neural parameters according to the control problem. 
As mentioned in Results, a lot of important parameters of the network are currently imposed to best fit the control case. Ideally, we would want the same network to be able to control different plants. This could be achieved by applying a discount factor on some parameters, which would allow us to adapt, for instance, the value of $f$ as the simulation progresses, finding the best time window to balance the number of spikes in input and the accuracy with which the target is traced. 
Finally, we can introduce learning on all the parameters of the network, to make our control as flexible as possible to a wide range of downstream systems. Not having to assume that we operate on completely observable plants would greatly expand the use cases of our algorithm, and would allow for testing of various neuroscience-derived hypotheses.

\section*{Materials and methods}

\subsection*{Simulation and network setup}
\paragraph*{Simulation details}
All the simulations in this paper were implemented in Python using the following packages. 
NumPy(version 1.23.5): for matrix operations and numerical computations. SciPy(version 1.10.1): for solving differential equations and simulating trajectories using \verb|solve_ivp|. Matplotlib(version 3.7.1) and Seaborn(version 0.13.2): for visualizing results, including raster plots, heatmaps, and system trajectories. Python control system library `control' (version 0.10.1): for the initializations of the continuous control example. 

\paragraph*{Code availability}
All simulations were run with Python 3.11.5. The
source code and all the necessary dependencies are available at https://gitlab.socsci.ru.nl/2023-phd-paolo-agliati/spiking-neurons-as-predictive-controllers-of-linear-systems.

\paragraph*{Time settings}
In each single SMD simulation (Figs.~\ref{fig4} to ~\ref{fig7}), the total time $T$ is set to $50$ seconds, the timestep $\delta t = 0.01 s$ divides each run into $5000$ timesteps $nT$. For the coupled SMD example (Fig.~\ref{fig8}) $T = 100s$, $\delta t = 0.01 s$, and $nT=10000$.

\paragraph*{Target settings}
For every simulation, the target $\mathbf{z}$ is only defined for the position component of the system's state, and it is initialized at zero. For Figs.~\ref{fig4}, ~\ref{fig6}, and ~\ref{fig7}, three target values are set: $z_{base} = 5$, from five second into the simulation, $z_{base} = 10$ after 15 seconds, and $z_{base} = 15$ after 30 seconds, and until the end of the simulation. For Fig.~\ref{fig8} we use similar $z_{base}$ steps, but we initialize them so each mass is attributed a different set of steps, which are progressively more positive or negative. In Fig.~\ref{fig5} we use bigger and irregular steps, to better show the activity of the four neurons: $z_{base} = -30$ from five second into the simulation, $z_{base} = 0$ after 15 seconds, $z_{base} = -25$ after 20 seconds, $z_{base} = 20$ after 30 seconds, and until the end of the simulation. 
Between each value of $z_{base}$, the parameter $z_{leak} = 0.5$ manages the exponential behavior of the target curve.

\paragraph*{Neurons and spike mapping}
For most runs, the number of neurons $N$ is fixed to two, except for Fig.~\ref{fig5} where $N=4$. In the scaling up example of coupled oscillators (Fig.~\ref{fig8}), $N=500$.
$\mathbf{B}$ is kept to spike the velocity component of a fixed value of $2$ or $-2$ for most simulations (the values relative to the positions are kept at zero). One exception is the physically unconstrained cases of Fig.~\ref{fig4}, where $\mathbf{B}$ can also allow spikes to the position of the same amounts ($2$ and $-2$). For Fig.~\ref{fig5}, $\mathbf{B}$ values are sampled from a normal distribution, with the second column (velocity) normalized to have norm $1$. The resulting value is then doubled. The same normalized $\mathbf{B}$ values are quadrupled in the coupled oscillator simulation. In this two cases, the dimensions of matrix $\mathbf{B}$ are modified to allow each one of the $N=4$ and $N=500$ neurons to have their spikes mapped onto the system.

\paragraph*{Future time window and spike cost}
The future window $f$ is kept at $0.3$ for the predictive simulations in Figs.\ref{fig4}, \ref{fig5} and \ref{fig7}. When using reactive control, $f$ is set to zero. For Fig.~\ref{fig6}, $f$ values in the subsequent simulations are linearly spaced from $0.1$ to $1$ (included), with the fourth value rounded up at $0.25$. 
$\mu$ is fixed at $0.3$ for the simulations in Figs.\ref{fig4} and \ref{fig7}. For Fig.~\ref{fig5}, $\mu$ is set to $0.1$. In Fig.~\ref{fig6}, $\mu$ values in the subsequent simulations are linearly spaced from $0.02$ to $1$ (included), with the third value rounded up at $0.25$. 

\paragraph*{Initial state and system matrix}
The initial state of the simulation $\mathbf{x_0}$ (at $T=0$) is always kept at $[0,0]$.
The system matrix $\mathbf{A}$ is also kept the same for all the single-SMD simulations (since we control the same SMD). The first row is set at $[0, 0.5]$, and the second at $[-0.1, -0.1]$. For Fig~\ref{fig8} the variable $M=10$ is introduced to indicate the number of masses to control.
To test robustness, we silence 180 random neurons 30 seconds into the simulation, and other 180 after 70 seconds. In this example, $A$ is scaled up to the dimensionality of the coupled systems based on $M$ (and $N$), and coupling values $\gamma = -0.3$ are introduced to make the position and acceleration of the masses dependent on each other.

\paragraph*{Continuous and filtered-spiking parameters}
The examples of continuous and filtered-spiking control in Fig.~\ref{fig7}A,B largely use the same settings as the other single-SMD simulations. In the continuous control case, we introduce an LQR gain matrix $\mathbf{K}$ to compute the control signal, initialized using the \verb|control.lqr| function from the python control package. To initialize $\mathbf{K}$, we use the matrix $\mathbf{B}$ that maps the control signal onto the velocity of system, set as $[0, 0.25]$. We will also need a control cost $R = 0.001$ (analogous to $\mu$ in the spiking control case), and finally a $K \times K$ matrix $\mathbf{Q}$ to introduce a cost on the dimensions of the system (analogous to $\mathbf{C}$ in our spiking paradigm). The first row of $\mathbf{Q}$ is set at $[1, 0]$, while the second is $[0, 1]$. For our filtered spiking control we will need a matrix $\mathbf{D} = [1, -1]$ to map the filtered spike traces onto the network read-out. For our example of Fig.~\ref{fig7}B, we use a spiking cost $\mu_f = 0.1$, and a decay factor $\lambda=1$ for the filtered spike traces.

\subsection*{Derivation}
\subsubsection*{Network}
In our work, we derive a spiking neural network (SNNs) of leaky integrate-and-fire (LIF) neurons from control principles. A network of $N$
such neurons is described by their voltage dynamics
\begin{equation} \label{eq:genericnet}
\mathbf{\dot{V}}(t) = -\lambda\mathbf{V}(t) +\mathbf{Fx}(t) + \mathbf{\Omega s}(t)
,\end{equation}
where \begin{math} \mathbf{V}(t) \in \R ^N \end{math}are the neurons’ voltages, \begin{math}\lambda \end{math} determines the membrane leak time-constant, \begin{math} \mathbf{x}(t) \in \R ^K \end{math} are the inputs, \begin{math}\mathbf{F} \in \R ^{N \times K} \end{math} are the forward weights, \begin{math}\mathbf{\Omega} \in \R ^{N \times N} \end{math} are the recurrent
weights, and \begin{math} \mathbf{s}(t) \in \R ^N \end{math} are the neural spike trains.
A spike is generated whenever a neuron’s voltage crosses its threshold, \begin{math}T_i \end{math}, and a spike train is described
as a sum of Dirac delta functions, \begin{math} s_{i} (t) = \sum_{t_j} \delta(t - t_j) \end{math}. Whenever a neuron spikes, its voltage is reset to a resting potential, \begin{math}R_i \end{math}. We now will generally be leaving out the time notation $(t)$ for notional clarity, unless it's important for a specific derivation step.

\subsubsection*{Dynamical System}
We want to control a linear dynamical system. We consider a $K$-dimensional linear dynamical system, over which we want to directly exert control using spike events.
The system is described by the following equation:
\begin{equation} \label{eq:genericsys}
\mathbf{\dot{x}} = \mathbf{A}\mathbf{x} + \mathbf{B}\mathbf{s} 
,\end{equation}
where \begin{math} \mathbf{x} \in \R ^K \end{math} represents the state of the system (one value for each dimension of the system), \begin{math} \mathbf{A} \in \R ^{K \times K} \end{math} is a matrix which defines the transformations in \begin{math} \mathbf{x} \end{math} that give rise to the dynamics (for instance, an oscillatory behavior), \begin{math} \mathbf{B} \in \R ^{K \times N} \end{math} is a matrix that maps the spike events onto the system, and \begin{math} \mathbf{s} \in \R ^N \end{math} are the spike trains, which represent our control signal.
Since it's a linear system, we can solve it analytically to derive the state of the system in a future time step \begin{math}f\end{math}, starting from the current state $\mathbf{x}_0$.

\noindent\paragraph*{No spike condition}
In the absence of any spiking control signal, we obtain:
\begin{equation*}
\mathbf{x}_f = e^{\mathbf{A} f} \mathbf{x}_0 
,\end{equation*}
where $\mathbf{x}_0  = \mathbf{x}(t)$ is the current state of the system, \begin{math}\mathbf{x}_f = \mathbf{{x}}(t+f)\end{math} is the predicted state after \begin{math}f\end{math} seconds, and \begin{math} e^{\mathbf{A} f} \end{math} is the matrix exponential of \begin{math} \mathbf{A} \end{math} times the future observation \begin{math}f\end{math}.
We introduce the shorthand \begin{math} e^{\mathbf{A} f} = \mathbf{A}_f \end{math} for ease of writing, resulting in: 
\begin{equation*} 
\mathbf{x}_f = \mathbf{A}_f \mathbf{x}_0 
.\end{equation*}

\noindent\paragraph*{Neuron \begin{math}i\end{math} spikes condition}
We now consider the effect of a single neuron \begin{math}i\end{math} spiking on the current state of the system, and how that affects the state of the system \begin{math}f\end{math} seconds into the future. For the condition where neuron \begin{math}i\end{math} spikes, the spike vector \begin{math} \mathbf{s} \end{math} is zero for all elements and one for the element corresponding to the spiking neuron \begin{math}i\end{math}. The operation \begin{math} \mathbf{B} \mathbf{s} \end{math} in Eq.\ref{eq:genericsys} results in considering the \begin{math}i\end{math}th column of matrix \begin{math} \mathbf{B} \end{math}. Here we denote that with \begin{math}\mathbf{b}_i\end{math}.
We can write the predicted state of the system after a spike as:
\begin{equation*} 
\mathbf{x}_f = \mathbf{A}_f (\mathbf{x}_0  + \mathbf{b}_i)
,\end{equation*} 
where $\mathbf{b}_i$ is the $i$th column of the matrix $\mathbf{B}$.

\subsubsection*{Losses}
To control the system, we set a target state \begin{math} \mathbf{\mathbf{z}} \end{math} which holds one value to be reached for each dimension of the system. To quantify the distance from this target, we define a loss as the \begin{math} L^2 \end{math} norm of the difference between the target and the predicted state:

\begin{equation*}
L = ||\mathbf{\mathbf{z}} - \mathbf{x}_f||^2 + \mu \Gamma(\mathbf s) + \alpha ||\mathbf r ||^2,
\end{equation*} 
where $\mathbf{z} \in \mathbb{R}^K$ is the target state, \begin{math}\mu\end{math} is a parameter that scales the cost for each spike, $\Gamma(\int_{t}^{t+ \delta t} \mathbf s(t) \dd{t})$ counts the spikes within a short time window (in practice within a simulation timestep), $\mathbf s$ are the spike trains, $\alpha$ is a parameter that scales the cost for the neuron's activity, and  $\mathbf r$ are the neurons' filtered spike traces. 
In the most general case, we can scale the part of this loss that describes the distance to the target. We introduce the matrix $\mathbf{C}$, which represents a cost on the dimensions of the system. This allows us to introduce a penalty on the values of the system's dimensions, which will weigh on the input to the controller.
The loss then becomes:
\begin{equation} \label{eq:weightedgenericloss}
L = (\mathbf{z} - \mathbf{x}_f)^T \mathbf{C} (\mathbf{z} - \mathbf{x}_f) + \mu \Gamma(\mathbf s) + \alpha ||\mathbf r ||^2 ,
\end{equation} 
where $\mathbf{C}\in \R ^{K \times K}$ is a cost matrix on each dimensions of the system. 
Assuming the perspective of a single neuron in our network, we can express this loss in the presence and absence of a spike.
\noindent\paragraph*{Loss with no spike}
In the absence of spikes, the loss of equation ~\eqref{eq:weightedgenericloss} becomes:
\begin{equation*} 
L^\text{ns} = (\mathbf{\mathbf{z}} - \mathbf{A}_f \mathbf{x}_0)^T \mathbf{C} (\mathbf{\mathbf{z}} - \mathbf{A}_f \mathbf{x}_0) + \alpha ||\mathbf r ||^2
.\end{equation*}
The error component in this loss can be renamed to $E$ for ease of writing, obtaining:
\begin{equation} \label{eq:Lns}
L^\text{ns} = E + \alpha \mathbf r^T \mathbf r
.\end{equation}
\noindent\paragraph*{Loss with neuron $i$ spike}
In our setup, neurons do not have access to the state of other neurons in the network, but rather compute a comparison between losses independently from each other. Thus, from the perspective of a single neuron \begin{math}i\end{math}, only one spike can be produced at the current time, and the function $\Gamma(\mathbf s)$ outputs a spike vector \begin{math}\mathbf{o}\end{math} which contains only zeros except for the \begin{math}i\end{math}th position, which is one. We name this vector $\mathbf{o}_i$. Under this assumption, we can rewrite $\Gamma(\mathbf s)$ as:
\begin{equation*}
\Gamma(\mathbf s) = \Gamma\left(\int_{t}^{t+ \delta t} \mathbf s(t) \dd{t}\right) = \sum_i^N (\mathbf{o}_i) = 1
,\end{equation*} 
the loss in Eq.~\eqref{eq:weightedgenericloss} then becomes:\begin{equation*} 
L_{i}^\text{s} = (\mathbf{\mathbf{z}} - \mathbf{A}_f (\mathbf{x}_0  + \mathbf{b}_i))^T \mathbf{C} (\mathbf{\mathbf{z}} - \mathbf{A}_f (\mathbf{x}_0  + \mathbf{b}_i)) + \mu + \alpha ||\mathbf r + \mathbf{o}_i ||^2
.\end{equation*} 
First, we can expand the error term of this loss, obtaining:
\begin{align*}
L_{i}^\text{s} = & E -2(\mathbf{A}_f \mathbf{b}_i)^T \mathbf{C} (\mathbf{\mathbf{z}} - \mathbf{A}_f \mathbf{x}_0) + \\&(\mathbf{A}_f \mathbf{b}_i)^T \mathbf{C} (\mathbf{A}_f \mathbf{b}_i) +\mu + \alpha ||\mathbf r + \mathbf{o}_i ||^2
.\end{align*} 
We now also expand the square term $\alpha ||\mathbf r + \mathbf{o}_i ||^2$, obtaining:
\begin{align*}
L_{i}^\text{s} = & E -2(\mathbf{A}_f \mathbf{b}_i)^T \mathbf{C} (\mathbf{\mathbf{z}} - \mathbf{A}_f \mathbf{x}_0) + \\
&(\mathbf{A}_f \mathbf{b}_i)^T \mathbf{C} (\mathbf{A}_f \mathbf{b}_i) + \mu + \alpha \mathbf{r}^T \mathbf{r} + 2\alpha \mathbf{o}_i{^T} \mathbf{r} + \alpha \mathbf{o}_i{^T}  \mathbf{o}_i
.\end{align*} 
Substituting from Eq.~\eqref{eq:Lns}:
\begin{align*}
L_{i}^\text{s} = & L^\text{ns}-2(\mathbf{A}_f \mathbf{b}_i)^T \mathbf{C} (\mathbf{\mathbf{z}} - \mathbf{A}_f \mathbf{x}_0) + \\&(\mathbf{A}_f \mathbf{b}_i)^T \mathbf{C} (\mathbf{A}_f \mathbf{b}_i) +\mu + 2\alpha \mathbf{o}_i{^T} \mathbf{r} + \alpha \mathbf{o}_i{^T}  \mathbf{o}_i
.\end{align*} 
As mentioned previously, $\mathbf{o}_i$ and \begin{math} \mathbf{o}_i{^T} \end{math} contain only zeros except for the $i$th position, which is one. This means \begin{math}  \mathbf{o}_i{^T}  \mathbf{o}_i = 1\end{math}. Let us consider the other term, \begin{math} \mathbf{o}_i{^T} \mathbf{r} \end{math}.
\begin{equation*}
\mathbf{o}_i{^T} \mathbf{r} = \sum \mathbf{o}_i \mathbf{r} = \mathbf{r}_i
\end{equation*}
Where $ \mathbf{r}_i$ indicates a vector containing only zeros except for the $i$th position, which holds the firing rates of neuron $i$.
We can write out loss with neuron $i$ spiking as: 
\begin{equation}\label{eq:Ls2}
\begin{split} 
L_{i}^\text{s} = &L^\text{ns} -2(\mathbf{A}_f \mathbf{b}_i)^T \mathbf{C} (\mathbf{\mathbf{z}} - \mathbf{A}_f \mathbf{x}_0) + \\&(\mathbf{A}_f \mathbf{b}_i)^T \mathbf{C} (\mathbf{A}_f \mathbf{b}_i) + \mu + \alpha (2\mathbf{r}_i+1)
.\end{split} 
\end{equation}

\subsubsection*{Neural Variables}

As mentioned in Discussion, previous works have treated spiking neurons as systems that can directly encode dynamical variables. Based on this view, they derived a ``greedy" spiking rule where a neuron $i$ fires whenever this results in a decrease of a cost function, which usually measures the distance between the dynamical variable $x$ and its estimate $\hat{x}$, or a target $z$. This is a fully local rule: a neuron's decision to fire depends solely on its current membrane potential and its threshold, which in turn depend on information locally accessible to the neuron at that moment. The neuron does not access global information about the activity of the network. This rule guarantees that each spike emitted by a neuron contributes to improving the representation (or the control) of the dynamical variable, and has therefore a clear interpretation within the computing objective of the network. We start deriving our network based on this simple firing condition:

\begin{equation*} 
L_{i}^\text{s}\leq L^\text{ns}
.\end{equation*}
Substituting from Eq.~\eqref{eq:Lns} and \eqref{eq:Ls2}:
\begin{align*}
&L^\text{ns} -2(\mathbf{A}_f \mathbf{b}_i)^T \mathbf{C} (\mathbf{\mathbf{z}} - \mathbf{A}_f \mathbf{x}_0) + (\mathbf{A}_f \mathbf{b}_i)^T \mathbf{C} (\mathbf{A}_f \mathbf{b}_i) \\&+ \mu + \alpha \mathbf{r}^T \mathbf{r} + \alpha (2\mathbf{r}_i+1) - L^\text{ns} \leq 0
,\end{align*} 
\begin{align*}
&-2(\mathbf{A}_f \mathbf{b}_i)^T \mathbf{C} (\mathbf{\mathbf{z}} - \mathbf{A}_f \mathbf{x}_0) + (\mathbf{A}_f \mathbf{b}_i)^T \mathbf{C} (\mathbf{A}_f \mathbf{b}_i) \\&+ \mu + \alpha (2\mathbf{r}_i+1)  \leq 0.
\end{align*}
Bringing the negative signed factors to the right side of the inequality, we obtain:
\begin{align*} 
&\mathbf{b}_i^T \mathbf{A}_f{^T} \mathbf{C}(\mathbf{\mathbf{z}} - \mathbf{A}_f \mathbf{x}_0) \geq \frac{1}{2}
\mathbf{b}_i^T \mathbf{A}_f^T \mathbf{C}\mathbf{A}_f \mathbf{b}_i \\& + \mu + \alpha(2\mathbf{r}_i+1))
\end{align*}
We can attribute the left side of this inequality to the threshold of the neuron $i$:
\begin{equation*} 
T_i = \frac{\mathbf{b}_i^T \mathbf{A}_f{^T} \mathbf{C} (\mathbf{A}_f \mathbf{b}_i)  + \mu + \alpha (2\mathbf{r}_i + 1)} {2}.
\end{equation*} 
As mentioned in Results, we showcase our simulations imposing an asynchronous firing mechanism on the network, so that only one neuron is allowed to fire in each timestep. However, our network derivation generalizes beyond this constraint. We introduce the cost term based on each neuron's firing rate in our losses ($\alpha ||\mathbf{r}||^2$ and $\alpha ||\mathbf{r} +\mathbf{s}_i||^2$), which is reflected on the neuron's threshold as $ \alpha (2\mathbf{r}_i + 1)$, and limits the neuron's ability to spike based on its own past activity. This spike-threshold adaptation mechanism allows us to retain the local aspect of our spiking rule, without causing too many control signals being inputted into the plant, and thus retaining the sparsity of the network's spiking regime. The spiking mechanism has very strong biophysical bases and has been present in previous modeling paradigms, especially in combination with LIF neuron models ~\cite{guoNeural2021, jonesStimulus2015}.

On the right side, all the time dependent factors can be attributed to the voltage of that neuron:
\begin{equation*} 
V_i = \mathbf{b}_i^T \mathbf{A}_f{^T} \mathbf{C}(\mathbf{\mathbf{z}} - \mathbf{A}_f \mathbf{x}_0).
\end{equation*}
This gives us a spiking condition for neuron $i$: \begin{math} {V_i} \geq T_i.\end{math}
If we do not pose any constraint on the dimensions of the system, we would make the matrix $\mathbf{C}$ an identity matrix $\mathbf{I}$, obtaining the voltage and threshold values described in Eq.~\eqref{eq:volnoc} and Eq.~\eqref{eq:thnoc}.
Stepping back from the single neuron perspective, and considering instead the voltages of all the neurons in the network, we obtain:
\begin{equation*}
\mathbf{V} = \mathbf{B}^T \mathbf{A}_f{^T} \mathbf{C}(\mathbf{\mathbf{z}} - \mathbf{A}_f \mathbf{x}_0).
\end{equation*}
From there, we pose:
\begin{equation*} 
\mathbf{G} = \mathbf{B}^T \mathbf{A}_f{^T} \mathbf{C}
,\end{equation*}
obtaining:
\begin{equation} \label{eq:V_i2}
\mathbf{V}= \mathbf{G} (\mathbf{\mathbf{z}}- \mathbf{A}_f \mathbf{x}_0).
\end{equation}
We now consider the rate of change of \begin{math}\mathbf{V} \end{math}, as well as the rate of change of all time-dependent variables:\\
\begin{equation} \label{eq:V_dot}
\dot{\mathbf{V}} = \mathbf{G}(\dot{\mathbf{\mathbf{z}}} - \mathbf{A}_f \dot{\mathbf{x}}_0)
,\end{equation}
the change of the current state of the system over time \begin{math} \dot{\mathbf{x}}_0 \end{math} can be expressed by Eq.~\eqref{eq:genericsys}:
\begin{equation*} 
\dot{\mathbf{x}}_0 = \mathbf{A}\mathbf{x}_0 + \mathbf{Bs}
.\end{equation*}
Substituting this into Eq.~\eqref{eq:V_dot} yields:
\begin{align}
\dot{\mathbf{V}} 
&= \mathbf{G} (\dot{\mathbf{z}} - \mathbf{A}_f (\mathbf{A} \mathbf{x}_0  + \mathbf{Bs})) \notag\\
&= \mathbf{G} (\dot{\mathbf{z}}  - \mathbf{A}_f \mathbf{A} \mathbf{x}_0  + \mathbf{A}_f \mathbf{Bs}) \notag\\
&= \mathbf{G} \dot{\mathbf{z}}  - \mathbf{G} \mathbf{A}_f \mathbf{A} \mathbf{x}_0   - \mathbf{G} \mathbf{A}_f \mathbf{Bs}
.\end{align}
We then add and subtract a \begin{math} \mathbf{V} \end{math} on the right side of the equation. We leave the $-\mathbf{V}$ as is, while we substitute the $+\mathbf{V}$ using the definition of \begin{math} \mathbf{V} \end{math} from Eq.~\eqref{eq:V_i2}, obtaining:
\begin{equation*}
\dot{\mathbf{V}} = -\mathbf{V} + \mathbf{G} \dot{\mathbf{z}} - \mathbf{G} \mathbf{A}_f \mathbf{A} \mathbf{x}_0  + \mathbf{G} \mathbf{z} - \mathbf{G} \mathbf{A}_f \mathbf{x}_0  -  \mathbf{G} \mathbf{A}_f \mathbf{Bs}
;\end{equation*}
which can be written as
\begin{equation*}
\dot{\mathbf{V}} = -\mathbf{V} + \mathbf{G} (\dot{\mathbf{z}} + \mathbf{z}) - \mathbf{G} \mathbf{A}_f \mathbf{(A+I)} \mathbf{x}_0  - \mathbf{G} \mathbf{A}_f \mathbf{Bs}
.\end{equation*}
We can trace this back to the form of a recurrent network of integrate and fire neurons: 
\begin{equation} \label{eq:network}
\mathbf{\dot{V}} = -\mathbf{V} + \mathbf{G} (\dot{\mathbf{z}} + \mathbf{z}) - \mathbf{F} \mathbf{x}_0  -  \mathbf{\Omega s}
,\,\end{equation}
where 
\begin{math} \mathbf{F}=\mathbf{G} \mathbf{A}_f \mathbf{(A+I)}\end{math}, and \begin{math} \mathbf{\Omega}=\mathbf{G} \mathbf{A}_f \mathbf{B} = \mathbf{B}^T \mathbf{A}_f{^T} \mathbf{C} \mathbf{A}_f \mathbf{B}.\end{math}\\
This full network equation describes the changes in $\mathbf{V} \in \mathbb{R}^N$ , which represents the membrane potentials. Here, $\mathbf{G} \in \R ^{N \times K}$ are the target input weights: forward connections that encode for the target $\mathbf{z}$. Considering the changes in target value over time $\dot{\mathbf{z}}$ as well as the actual value of the target $\mathbf{z}$ allows for the network to trace rapid changes of the target. Only encoding for the current value of $\mathbf{z}$ would still allow to control the system, but with a slower tracing of the target. The matrix $\mathbf{F} \in \R ^{N \times K}$ represents the state input weights, that contain the forward connections to encode the current system's state $\mathbf{x_0}$, $\mathbf{\Omega} \in \R ^{N \times N}$ are the recurrent weights, and $\mathbf{s}$ are the spikes.
Starting from a greedy spiking condition, we obtained a network of leaky integrate-and-fire (LIF) neurons in the form of Eq.~\eqref{eq:genericnet}. Such network has a voltage decay term, forward weights that map the target onto the network, forward weights that encode the system state in input, recurrent connectivity, and spikes, and is able to control our linear dynamical system. For a $K$-dimensional system to control and a network of $N$ neurons, \begin{math} \mathbf{B} \end{math} is a $N \times K$ matrix, while \begin{math} \mathbf{A}_f \end{math} and $\mathbf{C}$ are $K \times K$ matrices. Thus, \begin{math} \mathbf{\Omega} = \mathbf{B}^T \mathbf{A}_f{^T}  \mathbf{C} \mathbf{A}_f \mathbf{B} \end{math} is of rank $K$. Hence, our spiking control solution uses a network with low-rank connectivity (low-rank with respect to the system that is controlling). 

\subsection*{Relation to spike coding networks}
Works in SCNs use a very similar derivation as the one described in Materials and Methods. In these works, a read-out is usually defined in the following form:

\begin{equation} \label{eq:SCN1}
\mathbf y = \mathbf D \mathbf r,
\end{equation}

where $\mathbf D \in \mathbb{R}^{K \times N}$ contains the decoding or output weights of all neurons, and $\mathbf y \in \mathbb{R}^K$ is the $K$-dimensional network read-out (also sometimes seen as the network state). $\mathbf r \in \mathbb{R}^N$ represent some form of filtered spike trains, ranging from simple exponential filtering (such that $\dot{\mathbf r} = -\lambda \mathbf r + \mathbf s$), or more complex post-synaptic response kernels. In both cases a loss is optimized such that the output of the network progressively tracks its input, $\mathbf x$. A loss between these two quantities can be defined:

\begin{equation} \label{eq:SCN2}
L = ||\mathbf x - \mathbf D \mathbf r||^2.
\end{equation}

Much like in our derivation, from this loss voltage, dynamics are derived by considering the rule that spikes should only happen if $L(\text{neuron spikes}) < L(\text{no spikes})$, which we took inspiration from for the control case. For works that make use of complex kernels either some prediction of the future state was also employed ~\cite{schwemmerConstructing2015}, or a delayed signal was used to explicitly model the difference between past and current state ~\cite{zeldenrustEfficient2021}. 

These derivations are largely similar to ours, with the crucial differences that we consider downstream linear systems of a general form, and that previous SCN works have mainly considered a single read-out given by filtered spike trains. In this sense, our work can be seen as a generalization of standard SCN paradigms (which exert `control' mainly over their own read-out), which expands the use of similar principles to explicitly control any downstream linear dynamical system.

\subsection*{Filtered spiking controller}

For the filtered spiking controller in Fig.~\ref{fig7}B, we used a more direct extension of SCNs. There, we consider a network that should produce a read-out which matches the optimal control expected from an LQR controller. This means our network spikes to constrain the loss between the network output, and the LQR control signal, such that:

\begin{equation} \label{eq:SCN3}
L = ||\mathbf u - \mathbf D \mathbf r||^2,
\end{equation}

where $\mathbf u = -\mathbf K (\mathbf x - \mathbf z)$ is the control signal as it would be generated by an LQR controller. When defining the LQR signal to approximate, we consider a $K$-dimensional system controlled by a $W$-dimensional control signal. The matrix $\mathbf{K} \in \mathbb{R}^{W \times K}$ is then the optimal feedback gain matrix can be found by solving an algebraic Riccati equation, given assumptions on the cost of state deviations and control actuation~\cite{bruntonDatadriven2019}.
This solution ensures that the control law $\mathbf u = -\mathbf K (\mathbf x - \mathbf z)$ optimally balances tracking performance and control cost when controlling the linear system.
Spiking activity that constrains the loss described in Eq.~\eqref{eq:SCN3} can be achieved by simply using a standard SCN network, while considering $-\mathbf K (\mathbf x - \mathbf z)$ as an input instead of just $\mathbf x$. 

\section*{Acknowledgments}

The authors thank Luc Selen and Yuzhen Qin for advice on the model's design and conceptualization.

\end{multicols}

\section*{References}
\printbibliography[heading=none]

@article{ahissarPerception2016,
  author = {Ahissar, Ehud and Assa, Eldad},
  title = {Perception as a Closed-Loop Convergence Process},
  journal = {eLife},
  year = {2016},
  month = may,
  volume = {5},
  pages = {e12830},
  doi = {10.7554/eLife.12830},
  publisher = {eLife Sciences Publications, Ltd},
  editor = {Kleinfeld, David},
}

@Webpage{ahmadvandNeuromorphic2023,
  title = {Neuromorphic Robust Framework for Concurrent Estimation and Control in Dynamical Systems Using Spiking Neural Networks},
  author = {Ahmadvand, Reza and Sharif, Sarah Safura and Banad, Yaser Mike},
  year = {2023},
  month = oct,
  publisher = {arXiv:2310.03873},
  type = {Preprint},
  eprint = {2310.03873},
  primaryclass = {cs},
  doi = {10.48550/arXiv.2310.03873},
  lastchecked = {2025-06-10},
  url = {https://arxiv.org/abs/2310.03873}
}

@book{andersonOptimal2007,
  title = {Optimal Control: Linear Quadratic Methods},
  shorttitle = {Optimal Control},
  author = {Anderson, Brian D. O. and Moore, John B.},
  year = {2007},
  month = feb,
  publisher = {Courier Corporation},
  googlebooks = {fW6TAwAAQBAJ},
  isbn = {978-0-486-45766-6},
  langid = {english}
}

@article{baratchartlocal2009,
  title = {On Local Linearization of Control Systems},
  author = {Baratchart, L. and Pomet, J.-B.},
  year = {2009},
  month = oct,
  journal = {Journal of Dynamical and Control Systems},
  volume = {15},
  number = {4},
  pages = {471--536},
  issn = {1573-8698},
  doi = {10.1007/s10883-009-9077-9},
  urldate = {2025-01-08},
  langid = {english}
}

@article{bazhenovRole2005,
  title = {Role of Network Dynamics in Shaping Spike Timing Reliability},
  author = {Bazhenov, Maxim and Rulkov, Nikolai F. and Fellous, Jean-Marc and Timofeev, Igor},
  year = {2005},
  month = oct,
  journal = {Physical Review E},
  volume = {72},
  number = {4},
  pages = {041903},
  publisher = {American Physical Society},
  doi = {10.1103/PhysRevE.72.041903},
  urldate = {2025-03-11}
}

@article{bernikernormative2019,
  title = {A Normative Approach to Neuromotor Control},
  author = {Berniker, Max and Penny, Steven},
  year = {2019},
  month = apr,
  journal = {Biological Cybernetics},
  volume = {113},
  number = {1-2},
  pages = {83--92},
  issn = {1432-0770},
  doi = {10.1007/s00422-018-0777-7},
  langid = {english},
  pmid = {30178151}
}

@article{boerlinPredictive2013,
  title = {Predictive Coding of Dynamical Variables in Balanced Spiking Networks},
  author = {Boerlin, Martin and Machens, Christian K. and Den{\`e}ve, Sophie},
  year = {2013},
  month = nov,
  journal = {PLOS Computational Biology},
  volume = {9},
  number = {11},
  pages = {e1003258},
  publisher = {Public Library of Science},
  issn = {1553-7358},
  doi = {10.1371/journal.pcbi.1003258},
  urldate = {2024-06-07},
  langid = {english}
}

@article{boerlinSpikebased2011,
  title = {Spike-Based Population Coding and Working Memory},
  author = {Boerlin, Martin and Den{\`e}ve, Sophie},
  year = {2011},
  month = feb,
  journal = {PLOS Computational Biology},
  volume = {7},
  number = {2},
  pages = {e1001080},
  publisher = {Public Library of Science},
  issn = {1553-7358},
  doi = {10.1371/journal.pcbi.1001080},
  urldate = {2025-03-18},
  langid = {english}
}

@article{brendelLearning2020,
  title = {Learning to Represent Signals Spike by Spike},
  author = {Brendel, Wieland and Bourdoukan, Ralph and Vertechi, Pietro and Machens, Christian K. and Den{\`e}ve, Sophie},
  year = {2020},
  month = mar,
  journal = {PLOS Computational Biology},
  volume = {16},
  number = {3},
  pages = {e1007692},
  publisher = {Public Library of Science},
  issn = {1553-7358},
  doi = {10.1371/journal.pcbi.1007692},
  urldate = {2025-03-11},
  langid = {english}
}

@book{bruntonDatadriven2019,
  title = {Data-Driven Science and Engineering: Machine Learning, Dynamical Systems, and Control},
  shorttitle = {Data-Driven Science and Engineering},
  author = {Brunton, Steven L. and Kutz, J. Nathan},
  year = {2019},
  month = jan,
  edition = {1},
  publisher = {Cambridge University Press},
  doi = {10.1017/9781108380690},
  urldate = {2025-07-06},
  copyright = {https://www.cambridge.org/core/terms},
  isbn = {978-1-108-38069-0},
  langid = {english}
}

@article{byadarhalymodular2012,
  title = {A Modular Neural Model of Motor Synergies},
  author = {Byadarhaly, Kiran V. and Perdoor, Mithun C. and Minai, Ali A.},
  year = {2012},
  month = aug,
  journal = {Neural Networks},
  volume = {32},
  pages = {96--108},
  issn = {08936080},
  doi = {10.1016/j.neunet.2012.02.003},
  urldate = {2025-06-10},
  copyright = {https://www.elsevier.com/tdm/userlicense/1.0/},
  langid = {english}
}

@article{calaimgeometry2022,
  title = {The Geometry of Robustness in Spiking Neural Networks},
  author = {Calaim, Nuno and Dehmelt, Florian A and Gonçalves, Pedro J and Machens, Christian K},
  editor = {Meister, Markus and Frank, Michael J and Doiron, Brent and Meister, Markus},
  year = {2022},
  month = may,
  journal = {eLife},
  volume = {11},
  pages = {e73276},
  publisher = {eLife Sciences Publications, Ltd},
  issn = {2050-084X},
  doi = {10.7554/eLife.73276},
  url = {https://doi.org/10.7554/eLife.73276},
  urldate = {2023-10-03}
}

@article{cisekcomputer1999,
  title = {Beyond the Computer Metaphor: Behavior as Interaction},
  author = {Cisek, Paul},
  journal = {Journal of Consciousness Studies},
  year = {1999},
  volume = 6,
  number = 11,
  pages = {125-142},
  langid = {english}
}

@article{deneveEfficient2016,
  title = {Efficient Codes and Balanced Networks},
  author = {Den{\`e}ve, Sophie and Machens, Christian K.},
  year = {2016},
  month = mar,
  journal = {Nature Neuroscience},
  volume = {19},
  number = {3},
  pages = {375--382},
  publisher = {Nature Publishing Group},
  issn = {1546-1726},
  doi = {10.1038/nn.4243},
  urldate = {2025-03-18},
  copyright = {2016 Springer Nature America, Inc.},
  langid = {english}
}

@article{eliasmithunified2005,
  title = {A Unified Approach to Building and Controlling Spiking Attractor Networks},
  author = {Eliasmith, Chris},
  year = {2005},
  month = jun,
  journal = {Neural Computation},
  volume = {17},
  number = {6},
  pages = {1276--1314},
  issn = {0899-7667},
  doi = {10.1162/0899766053630332},
  langid = {english},
  pmid = {15901399}
}

@Webpage{eilersStability2025,
  title = {On the Stability of Event-Based Control with Neuronal Dynamics},
  author = {Eilers, Luke and Stapmanns, Jonas and Dias, Catarina and Pfister, Jean-Pascal},
  year = {2025},
  month = nov,
  publisher = {arXiv:2511.18015},
  type = {Preprint},
  eprint = {2511.18015},
  primaryclass = {eess},
  doi = {10.48550/arXiv.2511.18015},
  urldate = {2026-02-06},
  archiveprefix = {arXiv},
  url = {https://arxiv.org/abs/2511.18015}
}

@article{felgenhauerstability2003,
  title = {On Stability of Bang-Bang Type Controls},
  author = {Felgenhauer, Ursula},
  year = {2003},
  month = jan,
  journal = {SIAM Journal on Control and Optimization},
  volume = {41},
  number = {6},
  pages = {1843--1867},
  publisher = {{Society for Industrial and Applied Mathematics}},
  issn = {0363-0129},
  doi = {10.1137/S0363012901399271},
  urldate = {2025-06-03}
}

@article{gollischRapid2008,
  title = {Rapid Neural Coding in the Retina with Relative Spike Latencies},
  author = {Gollisch, Tim and Meister, Markus},
  year = {2008},
  month = feb,
  journal = {Science (New York, N.Y.)},
  volume = {319},
  number = {5866},
  pages = {1108--1111},
  issn = {1095-9203},
  doi = {10.1126/science.1149639},
  langid = {english},
  pmid = {18292344}
}

@article{guoNeural2021,
  title = {Neural Coding in Spiking Neural Networks: A Comparative Study for Robust Neuromorphic Systems},
  shorttitle = {Neural Coding in Spiking Neural Networks},
  author = {Guo, Wenzhe and Fouda, Mohammed E. and Eltawil, Ahmed M. and Salama, Khaled Nabil},
  year = {2021},
  month = mar,
  journal = {Frontiers in Neuroscience},
  volume = {15},
  publisher = {Frontiers},
  issn = {1662-453X},
  doi = {10.3389/fnins.2021.638474},
  urldate = {2025-03-18},
  langid = {english}
}

@incollection{guRobust2005,
  title = {Robust Control of a Mass-Damper-Spring System},
  booktitle = {Robust Control Design with MATLAB},
  author = {Gu, Da-Wei and Petkov, Petko Hristov and Konstantinov, Mihail Mihaylov},
  editor = {Gu, Da-Wei and Petkov, Petko Hristov and Konstantinov, Mihail Mihaylov},
  year = {2005},
  pages = {101--162},
  publisher = {Springer},
  address = {London},
  doi = {10.1007/1-84628-091-5-\_8},
  urldate = {2025-04-01},
  isbn = {978-1-84628-091-7},
  langid = {english}
}

@article{gutigspike2014,
  title = {To Spike, or When to Spike?},
  author = {G{\"u}tig, Robert},
  year = {2014},
  month = apr,
  journal = {Current Opinion in Neurobiology},
  series = {Theoretical and Computational Neuroscience},
  volume = {25},
  pages = {134--139},
  issn = {0959-4388},
  doi = {10.1016/j.conb.2014.01.004},
  urldate = {2025-03-18}
}

@inproceedings{huangDynamical2018,
  title = {Dynamical Spiking Networks for Distributed Control of Nonlinear Systems},
  booktitle = {2018 {{Annual American Control Conference}} ({{ACC}})},
  author = {Huang, Fuqiang and Ching, ShiNung},
  year = {2018},
  month = jun,
  pages = {1190--1195},
  issn = {2378-5861},
  doi = {10.23919/ACC.2018.8430996},
  urldate = {2024-06-07}
}

@inproceedings{huangOptimizing2017,
  title = {Optimizing the Dynamics of Spiking Networks for Decoding and Control},
  booktitle = {2017 {{American Control Conference}} ({{ACC}})},
  author = {Huang, Fuqiang and Riehl, James and Ching, ShiNung},
  year = {2017},
  month = may,
  pages = {2792--2798},
  issn = {2378-5861},
  doi = {10.23919/ACC.2017.7963374},
  urldate = {2025-04-11}
}

@article{jonesStimulus2015,
  title = {A Stimulus-Dependent Spike Threshold Is an Optimal Neural Coder},
  author = {Jones, Douglas L. and Johnson, Erik C. and Ratnam, Rama},
  year = {2015},
  month = jun,
  journal = {Frontiers in Computational Neuroscience},
  volume = {9},
  publisher = {Frontiers},
  issn = {1662-5188},
  doi = {10.3389/fncom.2015.00061},
  urldate = {2026-02-06},
  langid = {english}
}

@book{kochBiophysics2004,
  title = {Biophysics of Computation: Information Processing in Single Neurons},
  shorttitle = {Biophysics of Computation},
  author = {Koch, Christof},
  year = {2004},
  month = oct,
  publisher = {Oxford University Press},
  googlebooks = {MK5oAgAAQBAJ},
  isbn = {978-0-19-976055-8},
  langid = {english}
}

@article{liuSpiking2023,
  title = {Spiking Neural-Networks-Based Data-Driven Control},
  author = {Liu, Yuxiang and Pan, Wei},
  year = {2023},
  month = jan,
  journal = {Electronics},
  volume = {12},
  number = {2},
  pages = {310},
  publisher = {Multidisciplinary Digital Publishing Institute},
  issn = {2079-9292},
  doi = {10.3390/electronics12020310},
  urldate = {2025-03-18},
  copyright = {http://creativecommons.org/licenses/by/3.0/},
  langid = {english}
}

@article{mainenReliability1995,
  title = {Reliability of Spike Timing in Neocortical Neurons},
  author = {Mainen, Z. F. and Sejnowski, T. J.},
  year = {1995},
  month = jun,
  journal = {Science (New York, N.Y.)},
  volume = {268},
  number = {5216},
  pages = {1503--1506},
  issn = {0036-8075},
  doi = {10.1126/science.7770778},
  langid = {english},
  pmid = {7770778}
}

@article{mancooUnderstanding2020,
  title = {Understanding Spiking Networks through Convex Optimization},
  author = {Mancoo, Allan and Keemink, Sander W and Machens, Christian K},
  journal = {Advances in neural information processing systems},
  volume = {33},
  pages = {8824-8835},
  year = {2020},
  langid = {english}
}

@article{mooreneuron2024,
  title = {The Neuron as a Direct Data-Driven Controller},
  author = {Moore, Jason J. and Genkin, Alexander and Tournoy, Magnus and {Pughe-Sanford}, Joshua L. and {de Ruyter van Steveninck}, Rob R. and Chklovskii, Dmitri B.},
  year = {2024},
  month = jul,
  journal = {Proceedings of the National Academy of Sciences},
  volume = {121},
  number = {27},
  pages = {e2311893121},
  publisher = {Proceedings of the National Academy of Sciences},
  doi = {10.1073/pnas.2311893121},
  urldate = {2025-03-11}
}

@article{nardinNonlinear2021,
  title = {Nonlinear Computations in Spiking Neural Networks through Multiplicative Synapses},
  author = {Nardin, Michele and Phillips, James W. and Podlaski, William F. and Keemink, Sander W.},
  year = {2021},
  month = nov,
  journal = {Peer Community In Circuit Neuroscience},
  pages = {100003},
  doi = {10.24072/pci.cneuro.100003},
  urldate = {2023-10-03},
}

@article{ouyangImpulsive2020,
  title = {Impulsive Synchronization of Coupled Delayed Neural Networks with Actuator Saturation and Its Application to Image Encryption},
  author = {Ouyang, Deqiang and Shao, Jie and Jiang, Haijun and Nguang, Sing Kiong and Shen, Heng Tao},
  year = {2020},
  month = aug,
  journal = {Neural Networks},
  volume = {128},
  pages = {158--171},
  issn = {0893-6080},
  doi = {10.1016/j.neunet.2020.05.016},
  urldate = {2025-05-12}
}

@article{pazzagliaBalancing2025,
  title = {Balancing Central Control and Sensory Feedback Produces Adaptable and Robust Locomotor Patterns in a Spiking, Neuromechanical Model of the Salamander Spinal Cord},
  author = {Pazzaglia, Alessandro and Bicanski, Andrej and Ferrario, Andrea and Arreguit, Jonathan and Ryczko, Dimitri and Ijspeert, Auke},
  year = {2025},
  month = jan,
  journal = {PLOS Computational Biology},
  volume = {21},
  number = {1},
  pages = {e1012101},
  publisher = {Public Library of Science},
  issn = {1553-7358},
  doi = {10.1371/journal.pcbi.1012101},
  urldate = {2025-06-10},
  langid = {english}
}

@Webpage{petriAnalysis2024,
  title = {Analysis of a Simple Neuromorphic Controller for Linear Systems: A Hybrid Systems Perspective},
  shorttitle = {Analysis of a Simple Neuromorphic Controller for Linear Systems},
  author = {Petri, E. and Scheres, K. J. A. and Steur, E. and Heemels, W. P. M. H.},
  type = {Preprint},
  year = {2024},
  month = sep,
  number = {arXiv:2409.06353},
  eprint = {2409.06353},
  primaryclass = {cs, eess},
  publisher = {arXiv},
  doi = {10.48550/arXiv.2409.06353},
  urldate = {2025-01-08},
  url = {https://arxiv.org/abs/2409.06353},
  archiveprefix = {arXiv}
}

@article{podlaskiApproximating2024,
  title = {Approximating Nonlinear Functions with Latent Boundaries in Low-Rank Excitatory-Inhibitory Spiking Networks},
  author = {Podlaski, William F. and Machens, Christian K.},
  year = {2024},
  month = apr,
  journal = {Neural Computation},
  volume = {36},
  number = {5},
  pages = {803--857},
  issn = {1530-888X},
  doi = {10.1162/neco\_a\_01658},
  langid = {english},
  pmid = {38658028}
}

@article{polykretisspiking2022,
  title = {A Spiking Neural Network Mimics the Oculomotor System to Control a Biomimetic Robotic Head without Learning on a Neuromorphic Hardware},
  author = {Polykretis, Ioannis and Tang, Guangzhi and Balachandar, Praveenram and Michmizos, Konstantinos P.},
  year = {2022},
  month = may,
  journal = {IEEE Transactions on Medical Robotics and Bionics},
  volume = {4},
  number = {2},
  pages = {520--529},
  issn = {2576-3202},
  doi = {10.1109/TMRB.2022.3155278},
  urldate = {2025-03-18}
}

@inproceedings{qinAnalytical2024,
  title = {Analytical Characterization of Epileptic Dynamics in a Bistable System},
  booktitle = {2024 {{IEEE}} 63rd {{Conference}} on {{Decision}} and {{Control}} ({{CDC}})},
  author = {Qin, Yuzhen and {El-Gazzar}, Ahmed and Bassett, Danielle S. and Pasqualetti, Fabio and Van Gerven, Marcel},
  year = {2024},
  month = dec,
  pages = {583--588},
  issn = {2576-2370},
  doi = {10.1109/CDC56724.2024.10886408},
  urldate = {2025-06-24}
}

@inproceedings{raffObservers2007,
  title = {Observers with Impulsive Dynamical Behavior for Linear and Nonlinear Continuous-Time Systems},
  booktitle = {2007 46th {{IEEE Conference}} on {{Decision}} and {{Control}}},
  author = {Raff, Tobias and Allgower, Frank},
  year = {2007},
  month = dec,
  pages = {4287--4292},
  issn = {0191-2216},
  doi = {10.1109/CDC.2007.4434613},
  urldate = {2025-05-12}
}

@article{schwemmerConstructing2015,
  title = {Constructing Precisely Computing Networks with Biophysical Spiking Neurons},
  author = {Schwemmer, Michael A. and Fairhall, Adrienne L. and Den{\'e}ve, Sophie and {Shea-Brown}, Eric T.},
  year = {2015},
  month = jul,
  journal = {The Journal of Neuroscience: The Official Journal of the Society for Neuroscience},
  volume = {35},
  number = {28},
  pages = {10112--10134},
  issn = {1529-2401},
  doi = {10.1523/JNEUROSCI.4951-14.2015},
  langid = {english},
  pmcid = {PMC6605339},
  pmid = {26180189}
}

@article{seungHow1996,
  title = {How the Brain Keeps the Eyes\,Still},
  author = {Seung, H. S.},
  year = {1996},
  month = nov,
  journal = {Proceedings of the National Academy of Sciences},
  volume = {93},
  number = {23},
  pages = {13339--13344},
  publisher = {Proceedings of the National Academy of Sciences},
  doi = {10.1073/pnas.93.23.13339},
  urldate = {2025-03-18}
}

@article{slijkhuisClosedform2023,
  title = {Closed-Form Control with Spike Coding Networks},
  author = {Slijkhuis, Filip S and Keemink, Sander W and Lanillos, Pablo},
  year = {2023},
  journal = {IEEE Transactions on Cognitive and Developmental Systems},
  doi = {10.1109/TCDS.2023.3320251},
  langid = {english}
}

@article{srivastavaMotor2017,
  title = {Motor Control by Precisely Timed Spike Patterns},
  author = {Srivastava, Kyle H. and Holmes, Caroline M. and Vellema, Michiel and Pack, Andrea R. and Elemans, Coen P. H. and Nemenman, Ilya and Sober, Samuel J.},
  year = {2017},
  month = jan,
  journal = {Proceedings of the National Academy of Sciences},
  volume = {114},
  number = {5},
  pages = {1171--1176},
  publisher = {Proceedings of the National Academy of Sciences},
  doi = {10.1073/pnas.1611734114},
  urldate = {2025-03-18}
}

@inproceedings{stagstedneuromorphic2020,
  title = {Towards Neuromorphic Control: A Spiking Neural Network Based PID Controller for UAV},
  shorttitle = {Towards Neuromorphic Control},
  booktitle = {Robotics: Science and Systems 2020},
  author = {Stagsted, Rasmus and Vitale, Antonio and Binz, Jonas and Renner, Alpha and Bonde Larsen, Leon and Sandamirskaya, Yulia},
  year = {2020},
  month = jul,
  publisher = {RSS},
  address = {Virtual Conference},
  doi = {10.15607/rss.2020.xvi.074},
  urldate = {2025-03-18},
  copyright = {info:eu-repo/semantics/openAccess},
  isbn = {978-0-9923747-6-1},
  langid = {english}
}

@article{vincentPositioning1989,
  title = {Positioning and Active Damping of Spring-Mass Systems},
  author = {Vincent, T. L. and Joshi, S. P. and Lin, Yeong Ching},
  year = {1989},
  month = dec,
  journal = {Journal of Dynamic Systems, Measurement, and Control},
  volume = {111},
  number = {4},
  pages = {592--599},
  issn = {0022-0434},
  doi = {10.1115/1.3153099},
  urldate = {2025-06-10}
}

@article{wangImpulsive2014,
  title = {Impulsive Control and Synchronization of Nonlinear System with Impulse Time Window},
  author = {Wang, Xin and Li, Chuandong and Huang, Tingwen and Pan, Xiaoming},
  year = {2014},
  month = dec,
  journal = {Nonlinear Dynamics},
  volume = {78},
  number = {4},
  pages = {2837--2845},
  issn = {1573-269X},
  doi = {10.1007/s11071-014-1629-1},
  urldate = {2025-05-12},
  langid = {english}
}

@article{wolfeSparse2010,
  title = {Sparse and Powerful Cortical Spikes},
  author = {Wolfe, Jason and Houweling, Arthur R and Brecht, Michael},
  year = {2010},
  month = jun,
  journal = {Current Opinion in Neurobiology},
  series = {Sensory Systems},
  volume = {20},
  number = {3},
  pages = {306--312},
  issn = {0959-4388},
  doi = {10.1016/j.conb.2010.03.006},
  urldate = {2025-03-18}
}

@article{yangImpulsive1999,
  title = {Impulsive Control},
  author = {Yang, Tao},
  year = {1999},
  month = may,
  journal = {IEEE Transactions on Automatic Control},
  volume = {44},
  number = {5},
  pages = {1081--1083},
  issn = {1558-2523},
  doi = {10.1109/9.763234},
  urldate = {2024-06-07}
}

@article{yangStability2007,
  title = {Stability Analysis and Design of Impulsive Control Systems with Time Delay},
  author = {Yang, Zhichun and Xu, Daoyi},
  year = {2007},
  month = aug,
  journal = {IEEE Transactions on Automatic Control},
  volume = {52},
  number = {8},
  pages = {1448--1454},
  issn = {1558-2523},
  doi = {10.1109/TAC.2007.902748},
  urldate = {2025-05-12}
}

@article{yesildirekFeedback1995,
  title = {Feedback Linearization Using Neural Networks},
  author = {Ye{\c s}ildirek, A. and Lewis, F. L.},
  year = {1995},
  month = nov,
  journal = {Automatica},
  volume = {31},
  number = {11},
  pages = {1659--1664},
  issn = {0005-1098},
  doi = {10.1016/0005-1098(95)00078-B},
  urldate = {2025-04-16}
}

@article{zeldenrustEfficient2021,
  title = {Efficient and Robust Coding in Heterogeneous Recurrent Networks},
  author = {Zeldenrust, Fleur and Gutkin, Boris and Den{\'e}ve, Sophie},
  year = {2021},
  month = apr,
  journal = {PLOS Computational Biology},
  volume = {17},
  number = {4},
  pages = {e1008673},
  publisher = {Public Library of Science},
  issn = {1553-7358},
  doi = {10.1371/journal.pcbi.1008673},
  urldate = {2024-06-07},
  langid = {english}
}

@article{zhangRepresentation1996,
  title = {Representation of Spatial Orientation by the Intrinsic Dynamics of the Head-Direction Cell Ensemble: A Theory},
  shorttitle = {Representation of Spatial Orientation by the Intrinsic Dynamics of the Head-Direction Cell Ensemble},
  author = {Zhang, K.},
  year = {1996},
  month = mar,
  journal = {Journal of Neuroscience},
  volume = {16},
  number = {6},
  pages = {2112--2126},
  publisher = {Society for Neuroscience},
  issn = {0270-6474, 1529-2401},
  doi = {10.1523/JNEUROSCI.16-06-02112.1996},
  urldate = {2025-03-18},
  chapter = {Articles},
  copyright = {{\copyright} 1996 by Society for Neuroscience},
  langid = {english},
  pmid = {8604055}
}

\section*{Appendix}

\subsection*{Control with synchronous firing}
As mentioned in Results and Materials and Methods, we employ an asynchronous firing rule in all the examples of this paper, to showcase the simplest case of our spiking control. Of course, the assumption that only one neuron fires at any one time is generally not realistic. Furthermore, regardless of the level of biological detail in the network, asynchronous firing could not work for certain control tasks. For instance, if there exists any time delay between the system reaching a state and the neurons detecting such state, then multiple neurons would fire in response to the same loss comparison, and the system would be over-controlled, eventually missing the target. In a control problem with enough neurons and enough dimensions (where the mapping of the control signals greatly overlap), this would result in most of the neurons firing almost all the time, completely losing the sparsity of the spiking activity, as shown in Supplementary Fig.~\ref{figA}A. This happens because our network is fully local: neurons do not have access to the state of other neurons in the network via their connectivity, and thus compute their spiking decision independently from each other. Important to note that for simulations with zero delays that use very small timesteps, the asynchronous firing regime is naturally maintained, but for larger timesteps or in the presence of delays this is no longer feasible. The derivation of our network accounts for these issues and proposes an architecture that does not require impositions on the neuron's firing mechanism, while still retaining the local properties. To do so, we introduce a spike-threshold adaptation term (see Materials and Methods), which scales the threshold of each neuron based on its past activity. As shown in Supplementary Fig.~\ref{figA}B in the Appendix, employing a spike-threshold adaptation mechanism allows to recover control performance and a sparse spiking regime in case delays are present in the system (or in general, in the absence of asynchronous firing).

\begin{figure} [!h]
\includegraphics[width=\textwidth]{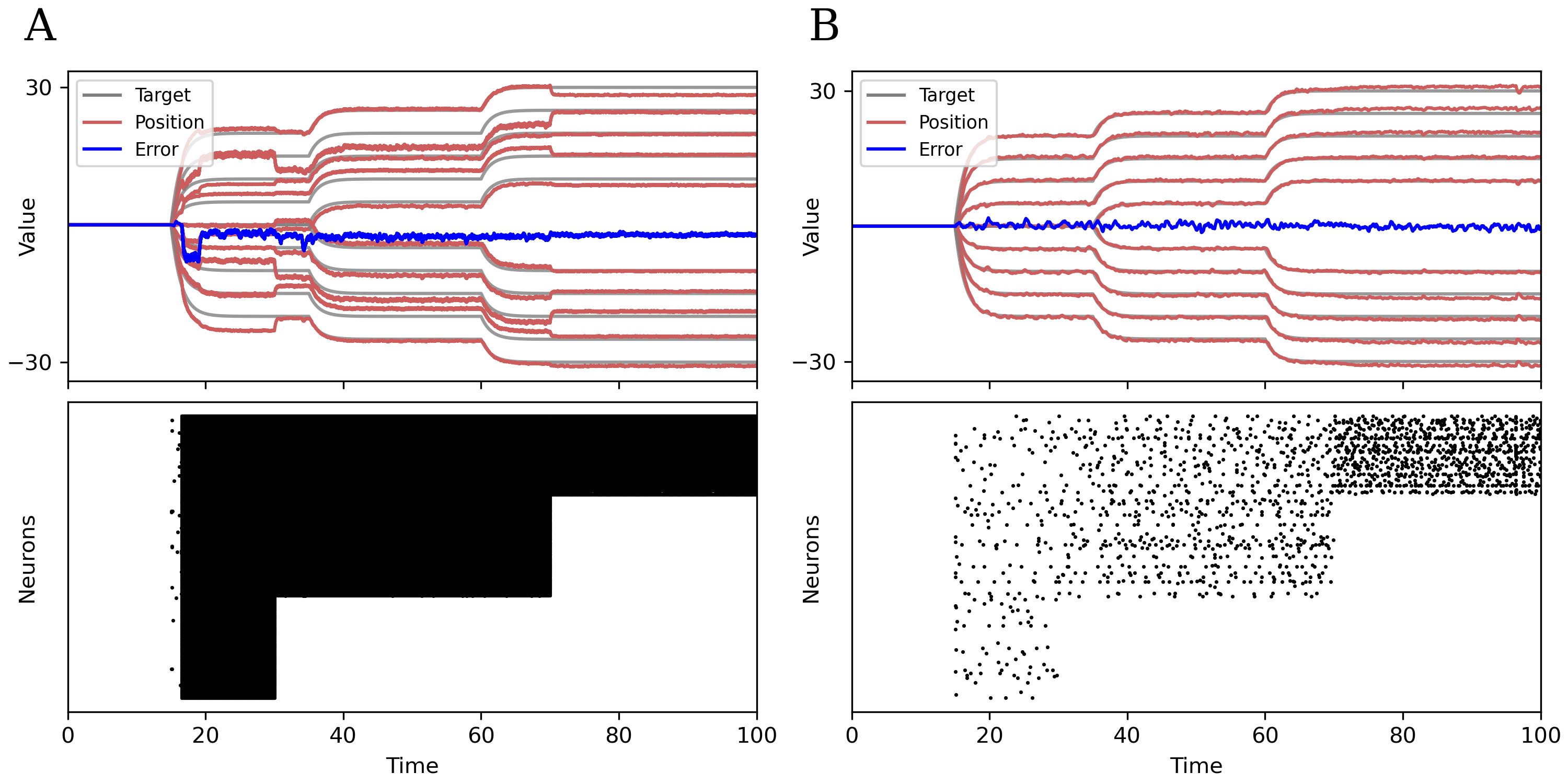}
\caption{Control examples without asynchronous firing.
A: Activity (top panel) and spike raster (lower panel) of a network of 500 neurons controlling a coupled SMD system of 10 oscillating masses. The example is identical to the one presented in Fig.~8, but the network showcased here does not operate with an asynchronous firing rule. In this case, more neurons are allowed to spike at the same timestep, and no spike-threshold adaptation mechanism is employed. As a result, the control performance is clearly affected, and all neurons are spiking at any given time. The local spiking rule of each neuron cannot account for this scenario, since no neuron has access to the state of other neurons in the network. B:  Activity (top panel) and spike raster (lower panel) of a network of 500 neurons controlling a coupled SMD system of 10 oscillating masses. The example is identical to the one presented in Fig.~8, but the network showcased here does not operate with an asynchronous firing rule. In this case, a spike-threshold adaptation mechanism is employed, so that neurons that had higher activity will be less likely to fire. As shown, this adaptive mechanism allows us to retain the local spiking rule without having an over-spiking regime, and without affecting the control performance.}
\label{figA}
\end{figure}

\subsection*{Energy measures: work done}

When considering the energy usage of our algorithm on the controlled plant (see Fig.~6), we refer to the effected acceleration onto the system. This quantity is very easily defined and interpreted, as it is only constituted by the change in velocity that occurs when a spike is emitted and mapped onto the system via the matrix $\mathbf{B}$. The energy we consider is thus the sum of the scaled spikes at their spiking times during the whole length of the simulation, $T_\text{end}$. The matrix $\mathbf B$ has zeros in all the rows and number $b_i$ for the row of neuron $i$, to map each spike $j$ onto the system.

$$W = \sum_{i=1}^N \sum_{j=1}^{K_i} \int_0^{T_\text{end}} b_i \delta (t - t^i_j) \dd{t} $$
where $N$ is the number of neurons and $K_i$ is the amount of spikes fired by neuron $i$, such that we sum over all neurons and all spikes.
Another principled way to consider the effect of the spiking control on the system state is to instead measure the work done on the system, by considering the kinetic energy before and after a neuron spikes.
Specifically, kinetic energy is computed as $E^{\text{kinetic}} = \frac{1}{2}m v(t)^2$ (considering a unit mass, $m=1$). Based upon the difference in kinetic energy before and after an instantaneous spike, the work done can be expressed as:
$$W = \sum_{i=1}^N \sum_{j=1}^{K_i} \lim_{\epsilon \rightarrow 0} \frac{1}{2} m (v(t^i_j + \epsilon)^2 - v(t^i_j - \epsilon)^2).$$
Effectively, this measures the change in kinetic energy before vs after the spike impacted the system.
Notably, when a spike occurs from neuron $i$, the change in velocity is specifically $b_i$, such that
$$W = \sum_{i=1}^N \sum_{j=1}^{K_i} \lim_{\epsilon \rightarrow 0} \frac{1}{2} m ((v(t^i_j - \epsilon) + b_i)^2 - v(t^i_j - \epsilon)^2).$$
In Supplementary Fig.~\ref{figB}, we show measures of energy input in the three control paradigms (continuous control, filtered spiking control, and spiking control) for the example SMD control of Fig.~7. We compare both the total work done on the system (Supplementary Fig.~\ref{figB}A) and the total number of spikes in the two spiking methods (Supplementary Fig.~\ref{figB}B).
As shown, there is no universal energy advantage of our spiking algorithm from a control-theoretic perspective. Rather, our contribution lies primarily in the formulation of the spiking control scheme, in the sparsity of the resulting spiking activity, and in the interpretability of our model. Our paradigm offers solutions for neuroscience models where spikes need to be sparsely computing control signals (we included a simple example of this in Fig.~9) or for any control applications where an impulsive control signal is needed.

\begin{figure} [!ht]
\centering
\includegraphics[width=0.9\textwidth]{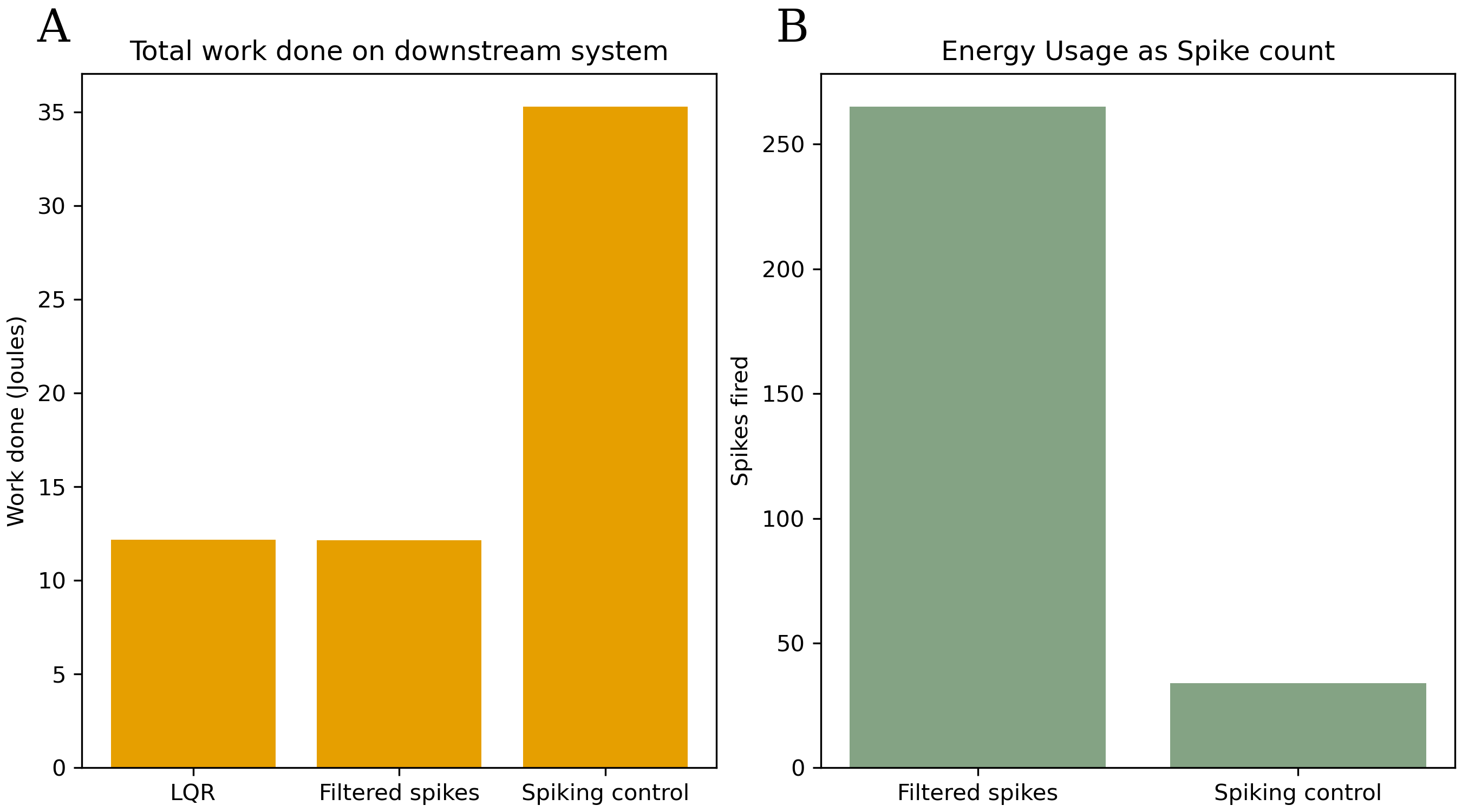}
\caption{Control input measures. A: Measures of work done in continuous, filtered spikes, and spiking control for a simple control task of a 2D linear SMD. B: Spike count for the same task for filtered spikes and spiking control.}
\label{figB}
\end{figure}

\end{document}